%% file: ms_gdwarf_MDF_arxiv.tex
\documentclass[twocolumn]{aastex62}
\usepackage{amsmath}
\received{November 25, 2021}
\revised{March 7, 2022}
\accepted{March 11, 2022}
\submitjournal{ApJ}
\shorttitle{MDF of elliptical galaxies}
\shortauthors{Choi et al.}
\input{defs.tex}

\begin{document}
	
\title{The Metallicity Distribution Function in Outer Halo Fields of Simulated Elliptical Galaxies \\
Compared to Observations of NGC 5128}

\correspondingauthor{Ena Choi}
\email{enachoi@kias.re.kr}
\author[0000-0002-8131-6378]{Ena Choi}
\affil{Quantum Universe Center, 
    Korea Institute for Advanced Study, 
    85 Hoegiro, Dongdaemun-gu, Seoul 02455, Republic of Korea; enachoi@kias.re.kr} 
     
\author{Jeremiah P. Ostriker}
\affil{Department of Astronomy, Columbia University, 
    550 West 120th Street, 
    New York, NY 10027, USA}
\affil{Department of Astrophysical Sciences, Princeton University, 
    Princeton, NJ 08544, USA}     

\author{Michaela Hirschmann}
\affil{Institut de Physique, Laboratoire d'astrophysique, \'Ecole Polytechnique F\'ed\'erale de Lausanne (EPFL), CH-1290 Versoix, Switzerland}
\affil{INAF - Osservatorio Astronomico di Trieste, via G. B. Tiepolo 11, I-34143 Trieste, Italy}
\affil{DARK, Niels Bohr Institute, University of Copenhagen, Lyngbyvej 2, DK-2100 Copenhagen, Denmark}
     
\author{Rachel S. Somerville}
\affil{Center for Computational Astrophysics, Flatiron Institute, 162 5th Ave., 
    New York, NY 10010, USA}
  
\author{Thorsten Naab}
\affil{Max-Planck-Institut f\"ur Astrophysik,
    Karl-Schwarzschild-Strasse 1, 85741 Garching, Germany}

\begin{abstract}
Stellar metallicity distribution functions (MDF) have been measured for resolved stellar populations in the outer halos of many galaxies in nearby groups. Among them, the MDF of NGC 5128, the central giant elliptical in the Centaurus group, provides essential constraints for theories of massive galaxy formation and hierarchical assembly. To investigate the formation and chemical evolution history of the outer halo of giant elliptical galaxies, we examine the chemical properties of three zoom-in high resolution cosmological hydrodynamical simulations of an NGC 5128-like giant elliptical galaxy and compare their outer halo MDFs to the observed one of NGC 5128. Even though the simulated galaxies have different merging histories and age distributions, all predicted MDFs are in good qualitative agreement with the observed one. The median metallicity of the simulated galaxies is on average $\rm [M/H]=-0.41 \pm 0.06$ compared to the observed value of $\rm [M/H]=-0.38 \pm 0.02$ for NGC 5128, and the dispersion in metallicity is $\sim 0.77$ dex for both observed and simulated galaxies. We investigate the origin of the stars ending up in the outer halo field of simulated galaxies and show that most have an `accreted' origin, formed in other small galaxies and later accreted in mergers. Only $\sim 15$ percent of the stars are formed `in situ' within the main progenitor of galaxy and radially migrate outwards. We show that the contribution of metal-rich in situ stars is sub-dominant in the outer halos of our simulated galaxies, but can be prominent in the inner regions. 
\end{abstract}

\section{Introduction}\label{intro}
Metallicity distribution functions (MDFs) have long been used to inform chemical evolution models, as they provide constraints on the star formation history of a galaxy and the merging history of its progenitors \citep[e.g.][]{Tinsley1980,Pagel1989}. The MDF has only been well characterized observationally in the solar neighborhood \citep{VandenBergh1962}. However, starting with the pioneering spectroscopic study on stars in Baade's window \citep{Rich1988}, the estimates of the MDF in several fields in the Milky Way have provided unique opportunities to understand the complex formation history of the Galaxy \citep[e.g.][]{Zoccali2008,Hill2011,Zoccali2017}. Recently, the Apache Point Observatory Galactic Evolution Experiment (APOGEE) survey provided MDFs across an unprecedented volume of the Milky Way, allowing a comprehensive characterization of the chemical history of the Galaxy \citep{Anders2014a,Hayden2015,Rojas-Arriagada2020}. These observed MDFs have been compared to predicted MDFs from cosmological hydrodynamical simulations and semi-analytic models of the Galaxy and provided new vital insights into the formation histories of the Galaxy \citep{Calura2012,Toyouchi2018,Mackereth2019}. 

For external galaxies, the study of the MDF is limited to galaxies in the Local Group and nearby groups. MDFs can be measured by analyzing the color distribution of stars on the red giant branch (RGBs) \citep[e.g.][]{Harris1998}. Deep imaging by the Hubble Space Telescope (HST) allows us to perform accurate stellar photometry in very crowded fields and has provided large photometric samples of individual RGB stars in nearby galaxies \citep[e.g.][]{Helmi2006,Harris2007,Harris2007a,Durrell2010,Mould2010,Monachesi2016,Cohen2020}. This opened up new windows into galaxy formation and assembly. By using observed MDFs of RGB stars as a constraint, detailed models of chemical evolution have been developed and applied to Local Group galaxies \citep{Calura2009,Tsujimoto2011,Romano2013,Homma2015}. The predicted MDFs from cosmological hydrodynamical simulations have been analyzed and intensively compared to those observed for dwarf galaxies \citep{Pilkington2012,Escala2018}, Milky Way-like disk galaxies \citep{Calura2012}, and ultrafaint dwarfs \citep{Jeon2017}.

However, there have been few if any studies of the MDF of massive elliptical galaxies formed in high-resolution cosmological simulations that would enable a comparison to current observations \citep[c.f. see][for non-cosmological simulations]{Bekki2002}. This may be partly due to the limited number of observed MDFs for resolved stars in elliptical galaxies, as resolved stellar photometry can only be applied to a small number of nearby systems. 

Among them, NGC 5128, also known as Centaurus~A, located at a distance of $\sim3.8$ Mpc \citep{Hui1993,Harris2009}, is currently the only viable target for wide-field resolved stellar population analysis of a giant elliptical (gE) galaxy. It is the dominant galaxy in the Cen A group and the closest gE galaxy in the nearby universe \citep{VandenBergh1976}. Despite its prominent dust lane along the photometric minor axis, the nature of its main elliptical component was well established via absorption lines and the surface brightness distribution \citep{Israel1998}. Its old red giant stars in the outer halo can be studied with HST imaging \citep{Soria1996,Harris1998}, or with a ground-based telescope under very good seeing conditions \citep{Crnojevic2013}. As HST observations reach at least $1-1.5$ mag below the RGB tip, resolved stars in several fields of the outer halo of NGC 5128 have been intensively investigated by many authors \citep{Harris1998,Harris2002,Rejkuba2005}, reaching out to a projected radius of 140 kpc \citep{Rejkuba2014}. Outermost fields are usually poorly sampled because of small number statistics, but several inner target fields are well sampled. For example, the target field of \citet{Rejkuba2005} with the galactocentric distance of $r_{\rm gc} = 40$ kpc in projection is well sampled with $\sim 2900$ stars. \citet{Rejkuba2005} found that the metallicity distribution of upper RGB stars in this outer halo field is moderately metal-rich and broad. 

The observed MDF in the outskirts of a gE can provide a significant constraint on models of elliptical galaxy formation. For example, in the currently favored `two-phase formation' model of massive galaxies \citep[e.g.][]{Oser2010}, intensive dissipational processes such as cold accretion \citep{Dekel2009} or gas-rich major mergers form an initially compact inner component, and then the outer part grows by the repeated merging of smaller galaxies through non-dissipational processes such as dry minor mergers \citep{Naab2007,Oser2010,2017ARA&amp;A..55...59N}. As this build-up of the stellar envelope is dominated by the accretion of old stars from lower mass galaxies, the model predicts that the outskirts of galaxies should be populated by stars with older ages and lower metallicities \citep{Hirschmann2015}, in contrast with other formation models such as monolithic collapse \citep{Eggen1962,Larson1975} or binary major mergers \citep{Toomre1972}. 

The properties of stars beyond the central regions of gE have mostly been explored with integrated light analyses due to the faintness of the outer regions \citep{Coccato2010,Greene2012}. Detailed information on the metallicity distribution of outer halo stars from the resolved stellar photometry can provide important constraints for massive galaxy formation. Comparing observed distribution functions of metallicity and age to the predictions of cosmological simulations of massive galaxy formation would allow us to test the two-phase massive galaxy formation scenario and better understand the complex formation histories of massive galaxies.

In this paper, we examine qualitatively and quantitatively how the simulated galaxies compare with the latest observations of the MDF of the resolved stellar populations in the local elliptical galaxy NGC 5128 in a full cosmological context. We first focus on the metallicity distribution of stars in the outer halo field with a galactocentric radius of $r_{\rm proj} \sim 40$~kpc, where we have well-sampled observational data from \citet{Rejkuba2005} for the comparison. We also analyze the build-up of the simulated MDF, tracing the formation history of stars in the outer halo. We then study the variations in the MDF of simulated galaxies with radial distance.

\section{Simulation and Method}\label{sec:method}
\input{tab1}
\subsection{Simulation}\label{subsec:simulation}
We use the suite of cosmological hydrodynamical simulations presented in \citet{Choi2017}. In the following, we provide a brief summary of the simulations, and we refer the reader to \citet{Choi2017} for further details. 

The cosmological zoom-in simulations of massive galaxies were generated with SPHGal \citep{2014MNRAS.443.1173H}, which is a modified version of the parallel smoothed particle hydrodynamics (SPH) code GADGET-3 \citep{2005MNRAS.364.1105S}. It includes a number of improvements to overcome the classical numerical fluid-mixing problems of SPH codes, such as a density-independent pressure-entropy SPH formulation \citep{2001MNRAS.323..743R}, an improved artificial viscosity \citep{2010MNRAS.408..669C}, and an artificial thermal conductivity \citep{2012MNRAS.422.3037R}. The initial conditions of zoom-in regions were adopted from \citet{Oser2010} who used cosmological parameters from WMAP3 \citep[][$h=0.72, \;\Omega_{\mathrm{b}}=0.044, \; \Omega_{\mathrm{dm}}=0.26, \;\Omega_{\Lambda}=0.74, \; \sigma_8=0.77 $, and $\mathrm{n_s}=0.95$]{2007ApJS..170..377S}. The simulations have a dark matter particle resolution of $m_{\mathrm{dm}} = 3.4 \times 10^{7} \Msun$, and a baryonic mass resolution of $m_{\mathrm{bar}}=5.8 \times 10^{6} \Msun$. The co-moving gravitational softening lengths are $\eps_{\mathrm{halo}} = 1.236 \rm \kpc$ for the dark matter and $\eps_{\mathrm{gas,star}} = 0.556 \rm \kpc $ for the gas and star particles.

The simulations include the star formation and chemical enrichment model of \cite{2013MNRAS.434.3142A}, which allows chemical enrichment by Type~I Supernovae (SNe), Type~II SNe, and asymptotic giant branch (AGB) stars. The chemical yields are respectively from \citet{1999ApJS..125..439I,1995ApJS..101..181W,2010MNRAS.403.1413K} for Type I, Type II SNe and AGB stars. The stellar kinetic feedback model is adopted from \citet{2017ApJ...836..204N}, in which winds from young massive stars, UV heating within \strom spheres of young stars, three-phase Supernova remnant input from both type I and type II SN feedback, and outflows from dying low-mass AGB stars are included. The simulations include the metal diffusion model from \cite{2013MNRAS.434.3142A}, in which the metal-enriched gas particles mix their metals with neighboring gas via turbulent diffusion of gas-phase metals.

In the simulations, new collisionless black hole particles with a mass of $10^5 \Msunh$ are seeded at the center of new emerging dark matter halos with mass above $1\times10^{11} \Msunh$. Then they can grow via mergers with other black holes and direct accretion of gas. The gas accretion onto a black hole is estimated with a Bondi-Hoyle-Lyttleton parameterization \citep{1939PCPS...34..405H,1944MNRAS.104..273B,1952MNRAS.112..195B}. The black hole mass accretion model also includes the soft Bondi criterion that statistically limits the accretion to the gas within the Bondi radius as in \cite{Choi2012a}. Super-Eddington accretion is allowed in the simulations as the accretion rate is not artificially capped at the Eddington rate. Instead, the Eddington force pushes electrons in gas particles radially away from the black hole, and the feedback processes summarized below quickly reduce the mass accretion. 

The simulations incorporate mechanical AGN feedback, which imparts mass and momentum to the surrounding gas. This model mimics the observed strong wind outflows \cite[e.g.][]{Arav2020} launched by radiation from radiatively efficient accretion onto a black hole \citep[e.g.][]{2000ApJ...543..686P,2004ApJ...616..688P}. 
The simulations also include Compton and photoionization heating and the associated radiation pressure effect of moderately hard X-ray radiation ($\sim 50$~keV) from the accreting black hole following \citet{2004MNRAS.347..144S,2005MNRAS.358..168S}. This mechanical and radiative AGN feedback model is shown to effectively suppress star formation in cosmological simulations of massive galaxies \citep{Choi2015a,Choi2017}.

In the simulations, the masses in 11 different species, H, He, C, N, O, Ne, Mg, Si, S, Ca, and Fe, are calculated explicitly for star and gas particles. The total metal abundance $Z$ is calculated using nine heavy elements and shown in units of Solar metallicity $Z_{\odot}$ = 0.0134 adopted from \citet{Asplund2009}.

The simulations reproduce the basic physical properties of observed massive elliptical galaxies. The most relevant observed global relation for this study is the stellar mass-metallicity relation (MMR), and the predicted stellar MMR of this simulation suite reproduces the overall shape and slope of the observed MMR \cite[see Figure 8 in][]{Choi2017}. \cite{Choi2020} also have shown that the simulations reproduce observed scaling relations between the iron abundance of hot gas halo and X-ray luminosity.

Of 30 galaxies in the \citet{Choi2017} simulation suite, we focus on three halos, m0163, m0125, and m0204, which have comparable masses and sizes to the local elliptical galaxy NGC 5128 that we focus on in this study. The physical quantities of these example galaxies are summarized in Table~\ref{tab:summary}.

\subsection{Measuring MDFs}\label{subsec:method}
In observations, the resolved stellar photometry is obtained only for a small target field. For example, the field of view of the Advanced Camera for Surveys (ACS) target field in \citet{Rejkuba2005} is $202''\times202''$, which corresponds to 3.7 kpc $\times$ 3.7 kpc at the 3.8 Mpc distance of NGC 5128 \citep{Harris2009}. Due to the resolution limit of our simulations, we did not use a small target field to select the outer halo stars as in the observations because it would contain too few star particles to obtain a meaningful measure of the MDF. Instead, we use all star particles within a projected distance of $\rm 37.5 < r_{\rm proj}/kpc < 42.5$, centered at $r_{\rm proj}=40$~kpc, which corresponds to that of \citet{Rejkuba2005} target field\footnote{The direction of the projection has a negligible effect on the MDF. We measured the MDF of the projected field along 100 randomly chosen directions relative to the main stellar body to test this. Changing the projection direction does not change the resulting MDF nor the median metallicity. The standard deviation of the median metallicity of stars sampled along 100 projections is 0.008. The resampling of stars within a larger region (e.g., $\rm 35 < r_{\rm proj}/kpc < 45$) also does not change the results.}. In addition, only stellar particles within the virial radius $r<r_{\rm vir}$ of the galaxies are selected to exclude foreground and background stars. The number of star particles used to obtain MDFs is 1100, 1531, and 715 for m0163, m125, and m0204, respectively. The number of RGB stars used to construct the observed MDF is 2058 \citep{Rejkuba2005}.

Since the limiting magnitude of the observations is $1.0 \-- 1.6$ mag below the tip of the RGB, metallicity distributions are only measured for the brightest RGB stars in many observational studies. These luminous red giant stars are biased against older populations, we correct this bias by applying a weight to each stellar particle based on its metallicity and age. Using the Yonsei-Yale model isochrones \citep{Yi2001,Demarque2004} and assuming a Kroupa IMF \citep{2001MNRAS.322..231K}, we calculate the weight, i.e., the fraction of observable mass above the observation limit to the total mass of a single stellar population (SSP). We adopted the limiting magnitude of $I \sim 29$ from \citet{Rejkuba2005}. The model weight is calculated for metallicities from $Z=10^{-5}$ to $0.08$, and for population ages from $t_{\rm age} =1$~Myr to $14$~Gyr. As the limiting magnitude of $I \sim 29$ is $1.0 \-- 1.6$ mag below the RGB tip, the observable mass fraction of RGB stars to the total mass of stellar population is higher for younger populations with $t_{\rm age} <1$~Gyr, and smaller for older stellar populations. In order to compare our results to observed MDFs, we apply this weight to each stellar particle when we tabulate the MDF for the simulated galaxies. Note that this correction does not dramatically alter the shape of the simulated MDFs since the outer halo region does not contain a significant population of young stars.

We take the metallicity of each star particle in the simulations to construct the MDF. Due to the limited resolution, we treat each star particle as a single stellar population. The age and metallicity of each star particle are used to calculate a weight and correct for the observational selection effect as described above. On the other hand, the metallicity value of each RGB star is obtained by interpolating between model isochrones in observations. A single age and a fixed $\alpha$-element abundance are usually adopted for model isochrones. The stellar metallicities are obtained in a fundamentally different fashion in observations and simulations. However, the discrepancy induced by different metallicity measurement methods in observations and simulations is expected to be minor in this study, as the outer halo stars are old with stellar age $t_{\rm age} \sim 10 $ Gyr in simulations, consistent with the commonly adopted stellar age of the isochrone fitting used in observations.

\begin{figure*}
\plotone{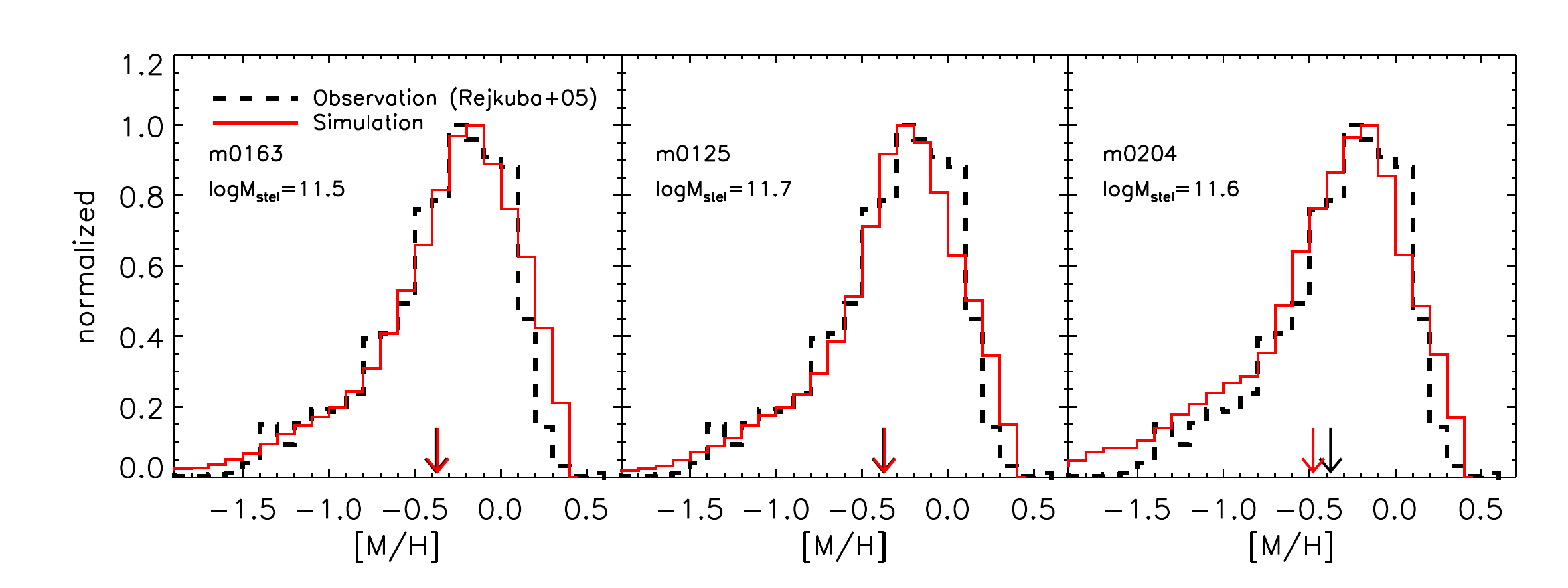}
\caption{Metallicity distribution functions (MDFs) of an outer halo field (within a projected radius of $37.5 < r_{\rm proj} < 42.5$ kpc) of the simulated galaxies m0163, m0125, and m0204, compared to the observed MDF of NGC 5128 from \citet{Rejkuba2005}, which is at $r_{\rm proj} \sim 40$ kpc. The downward arrows mark the median value of the stellar metallicity of each galaxy. The number of star particles used to obtain simulated MDFs is 1100, 1531, and 715 for m0163, m125, and m0204, respectively. The simulated MDFs include a smoothing, representing an estimated observational uncertainty of $\Delta {\rm [M/H]}=0.1$. There is excellent agreement between the simulation predictions and the observations. \label{fig:mdfobs} }
\end{figure*}

\begin{figure*}
\plotone{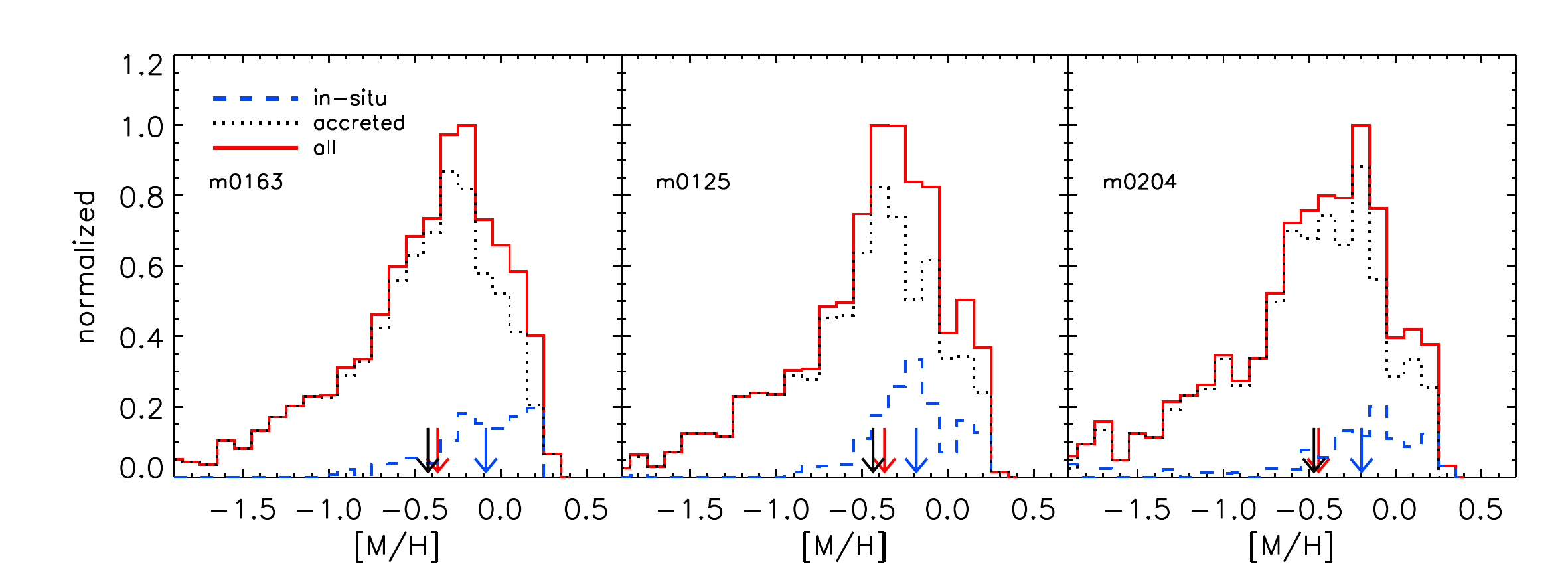}
\caption{MDF of all stars (solid red line) in the outer halo field of the simulated galaxies m0163, m0125, and m0204, and MDF of {\it in situ} formed stars (blue dashed), and MDF of those formed outside the galaxy and {\it accreted} later (black dotted). The downward arrows mark the median value of the stellar metallicity of each component. The outer halos of simulated massive elliptical galaxies mainly consists of accreted stars. \label{fig:mdftwophase} }
\end{figure*}

\begin{figure*}
 \plotone{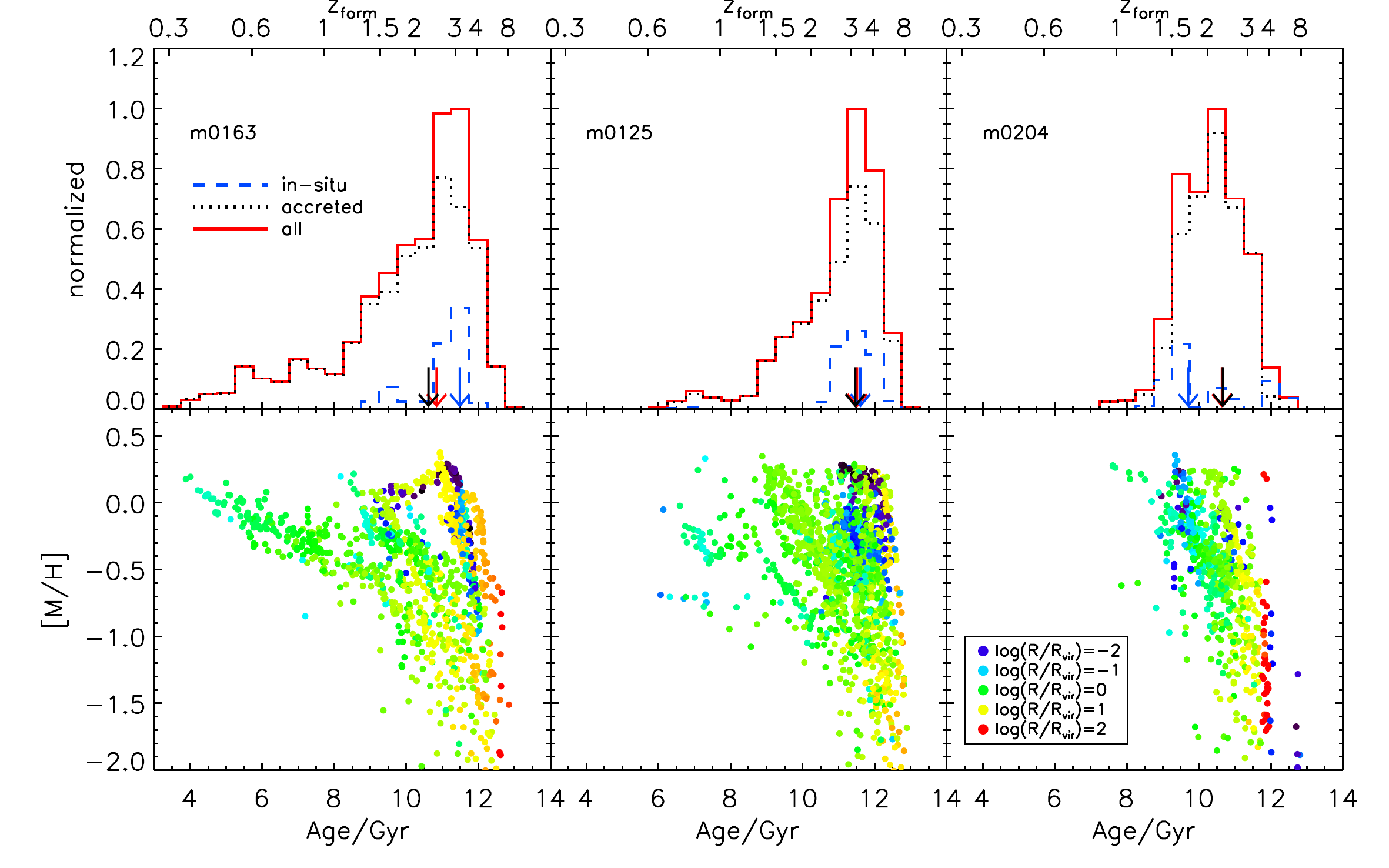}
 \caption{(Top) Histograms of stellar age for stars in the outer halo field of the simulated galaxies m0163, m0125, and m0204. The age distribution of all stars, {\it in situ} formed stars and {\it accreted} stars are shown by solid red lines, blue dashed lines, and black dotted lines, respectively. The downward arrows correspond to the median age of stars of each component. (Bottom) Age-metallicity relations (AMRs) of stars in the outer halo fields of simulated galaxies. Note the strong variations among the three different simulations. Color coding within each panel corresponds to the location where the star particle was formed, ranging from the inner region (dark blue/{\it in situ}) to the outer region (red/{\it ex situ}). \label{fig:agehist} }
\end{figure*}

\section{Results} \label{sec:result}
We first focus on the metallicity distribution of stars in the outer halo of galaxies, within a projected radius of $37.5 < r_{\rm proj} < 42.5$ kpc. The projected radius of this annular region corresponds to the galactocentric distance of the \cite{Rejkuba2005} observations of stars in NGC 5128 ($r_{\rm proj} \sim 40$ kpc). 

In Figure~\ref{fig:mdfobs}, we show MDFs of stars at $r_{\rm proj} \sim 40$ kpc for three example simulated galaxies. The observed MDF of an outer-halo field in NGC 5128 derived from the upper RGB photometry from \cite{Rejkuba2005} is also shown for comparison. The simulations can approximately reproduce the observed MDF, including its range of metallicity and the location of the peak. Both observed and simulated MDFs show a slow increase toward higher metallicities and a steeper decline after the peak. The MDFs of all simulated galaxies exhibit a peak at around $\rm [M/H]=-0.3$, similar to the observed peak. Compared to less massive local galaxies, the metallicity distribution of this giant E/S0 galaxy is broad and moderately metal-rich, with median metallicity $\rm [M/H]=-0.377$ \citep{Rejkuba2005}. The median value of the metallicity distributions of the simulated galaxies is $-0.367$, $-0.370$, and $-0.478$ for m0163, m0125, and m0204, respectively. The example galaxy m0204, with an effective radius $r_{\rm eff}$ a factor of two smaller than NGC 5128, is the most metal-poor and shows a more extended metal-poor tail.

Individual differences in MDFs of massive galaxies reflect a wide range of complex star formation and merger histories. To investigate the origin of stars that constitute the outer fields of massive galaxies, we traced the origin of each star particle included in the MDF of Figure~\ref{fig:mdfobs} and checked its formation location. For halo stars within a projected radius of $37.5 < r_{\rm proj} < 42.5$ kpc, we find that some are made in situ, within 10 \% of the virial radius ($r_{10} \equiv 0.1 \times r_{\rm vir}$) of the main halo, while most are formed ex situ, outside of $r_{10}$, and later accreted. In Figure~\ref{fig:mdftwophase} we compare the MDF of accreted stars and that of in situ stars for each example galaxy. We find that the outer halo fields of all example galaxies mainly consist of accreted stars. The shape and peak of the MDF of accreted stars are almost identical to that of all stars. On the other hand, in situ formed stars are mostly metal-rich with median metallicity $\rm [M/H] \sim -0.2$, and their MDFs lack the low metallicity tails. The median metallicity of in situ stars is about 0.3 dex higher than that of accreted stars.

\begin{figure*}
\gridline{\fig{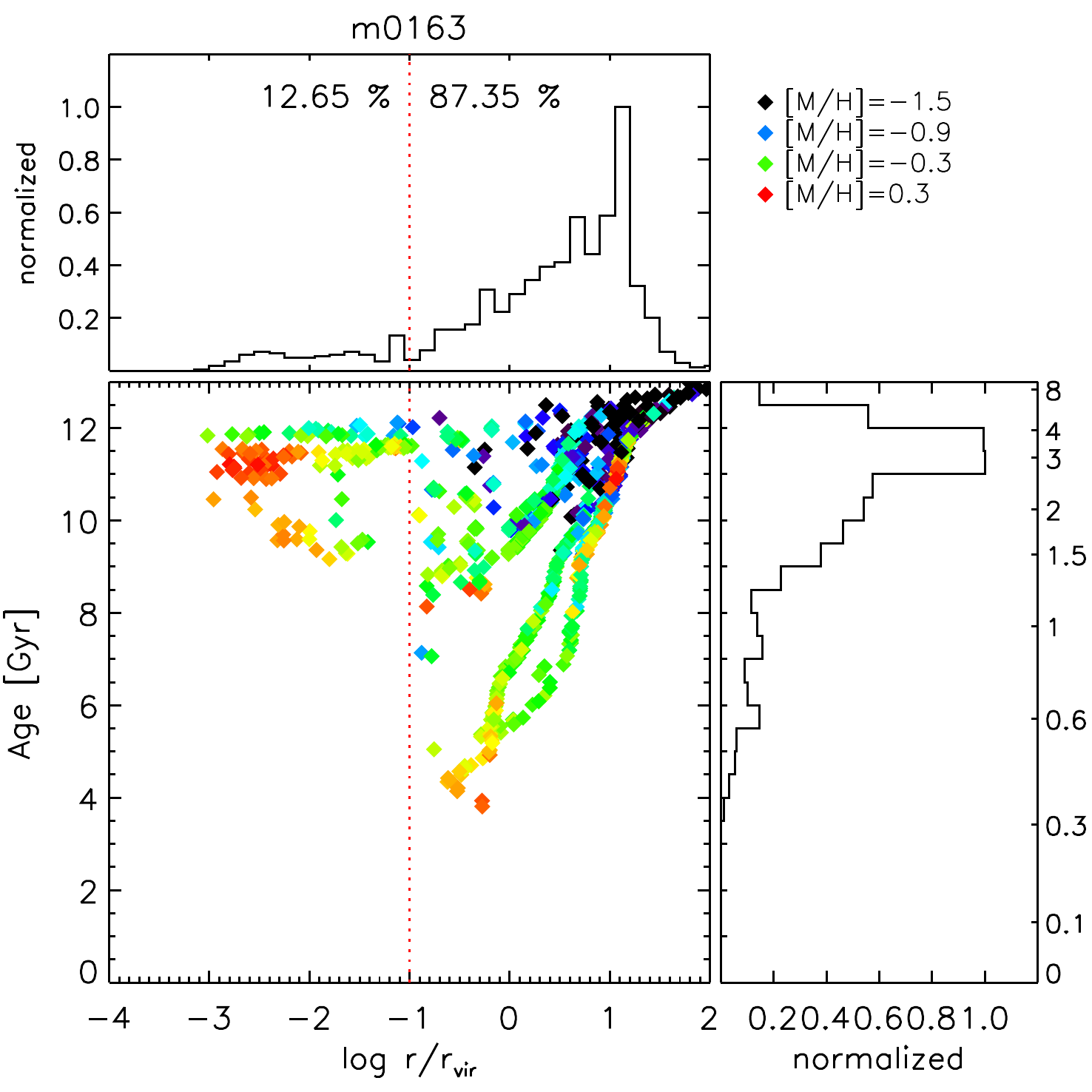}{0.34\textwidth}{(a)}
     \fig{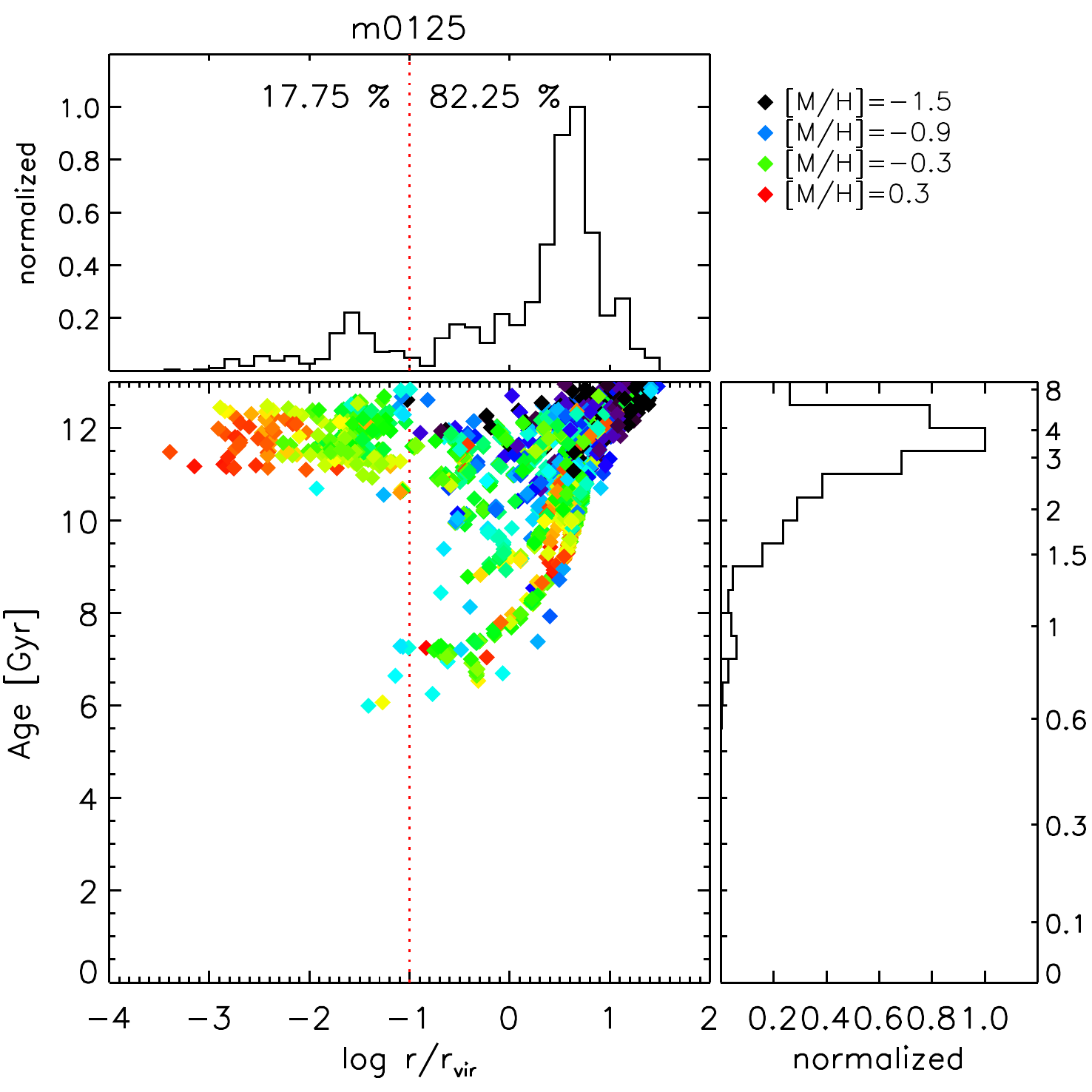}{0.34\textwidth}{(b)}
     \fig{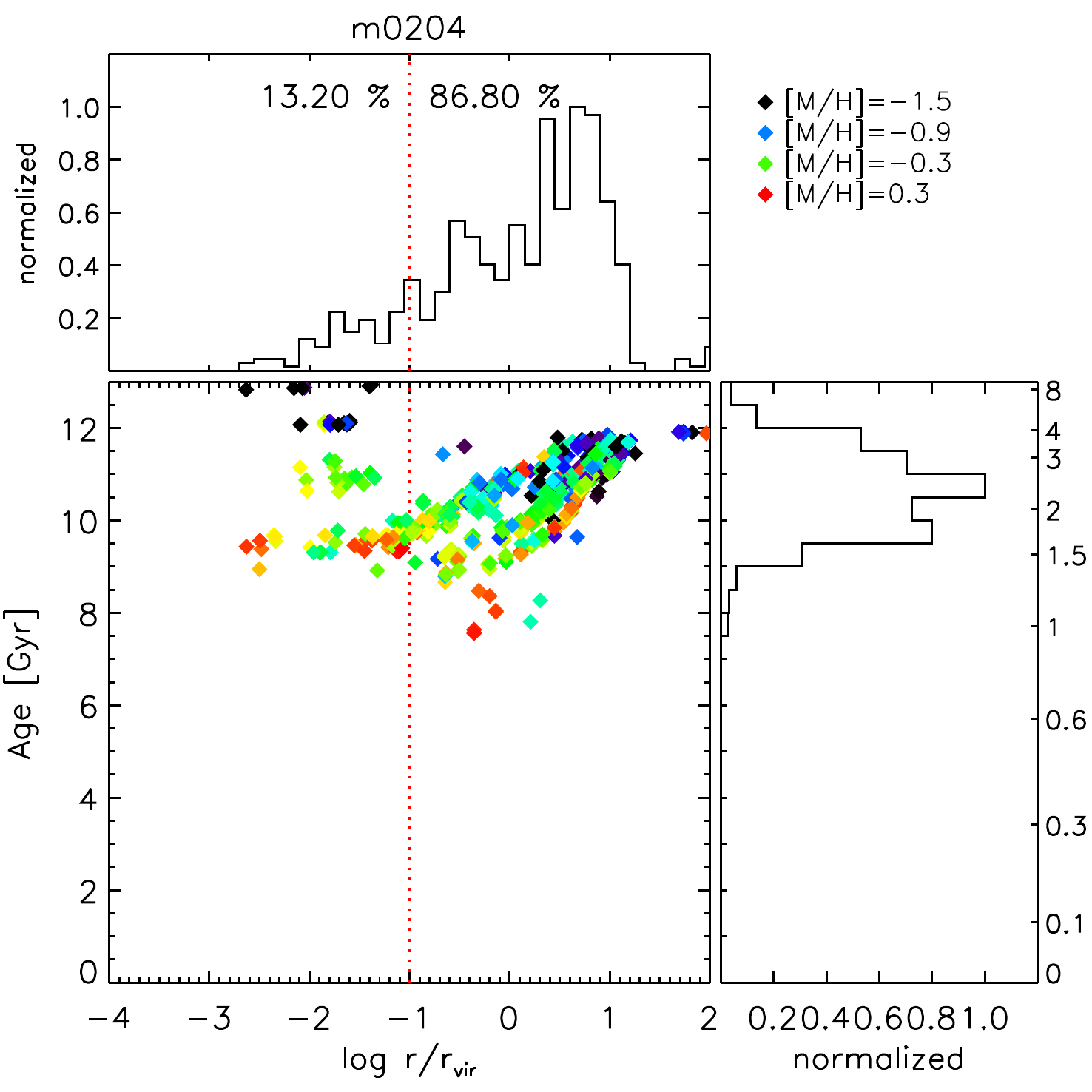}{0.34\textwidth}{(c)}}
 \caption{Stellar origin diagram for all stars in the outer halo field of example galaxies (a) m0163, (b) m0125, and (c) m0204. Each symbol marks the star particle's formation time and formation location in units of the virial radius of the main galaxy at that time. The red vertical dotted line indicates the 10 \% of the virial radius, which is commonly used to delineate the main stellar body of the galaxy. The metallicity of stars are color coded, ranging from $\rm [M/H]=-1.5$ (black/metal poor) to $\rm [M/H]=0.3$ (red/metal rich). Note that metal rich stars mostly have in -situ (central) origins, and metal poor stars have ex situ (external) origins. The histograms of formation location are shown in upper panels with the fraction of stars that form either inside $r_{10}$ (in situ) or outside $r_{10}$ (accreted). \label{fig:origin} }
\end{figure*}

The upper panels of Figure~\ref{fig:agehist} show the histogram of stellar ages for the simulated galaxies. The distributions of stellar ages of in situ and accreted stars are also shown respectively for comparison. Again, the stellar age distribution of outer halo field stars is dominated by the accreted stellar population. They are mostly old and exhibit a peak at $\sim 11.5 \rm~Gyr$, but there are a considerable number of young and intermediate-age stars. These young and intermediate-age stellar populations in the outer halo mainly have an accreted origin, while in situ formed stars are mostly old except for the case of the example galaxy m0204. Compared to the other two example galaxies, the stellar population of the galaxy m0204 is older, with a dearth of stars younger than $\sim \rm 7~Gyr$. Its in situ formed stellar population is younger compared to the other galaxies, with the median age of $t_{\rm age} \rm \sim 9.5~Gyr$.

To investigate the formation history of stars in the outer halo fields of simulated galaxies in detail, we show the age-metallicity relations (AMRs) of these stars in the lower panels of Figure~\ref{fig:agehist}. The color of each simulated star particle scales with its formation location in units of the virial radius of the main galaxy at that time. A striking feature of Figure~\ref{fig:agehist} is that the AMRs of the three example galaxies are quite different, especially for younger stars, even though their MDFs are similar (see Figure~\ref{fig:mdfobs}). This demonstrates that different elliptical galaxies march at different paces but reach very similar destinations.

The AMRs of the three example galaxies show similar trends for old stars. For all galaxies, the most metal-poor stars are the oldest. However, the oldest stars are not always the most metal-poor ones, as these galaxies have old stars that span a broad range in metallicity from $\rm -2.0 <[M/H]<0.3$. This indicates that the initial chemical enrichment was fast, already reaching solar or twice-solar metallicity for the $\rm \sim 11-12~Gyr$ old population. This fast chemical enrichment happens within the main progenitor ({\it in situ}, dark blue colored symbols) as well as within the smaller galaxies where the accreted stars are made ({\it ex situ}, orange-red colored symbols). These smaller systems are made at early times and contain primarily old stars, but we see that some of these systems also undergo a fast chemical enrichment at early times, adding metal-rich stellar populations to the final galaxy when they merge in.

On the other hand, the AMRs of three example galaxies show different features for the younger stellar population. First of all, the galaxy m0163 shows an extended and correlated AMR for $t_{\rm age} < 8~\rm Gyr$. This thin relation of young stars that predominantly formed at around $r\sim r_{\rm vir}$, is similar to those predicted by non-cosmological homogeneous galactic chemical evolution models \citep[e.g., see Figure 18 of][and references therein]{Carrera2008}. A period of slow metallicity evolution with small scatter until around $\rm 4~Gyr$ ago for accreted stars indicates that a small merging galaxy system underwent a period of star formation and chemical evolution. Secondly, the galaxy m0125 also shows a thin relation of young to intermediate age accreted stars as predicted by classical galactic chemical evolution models, but overall this feature is less apparent with a lack of young stars with $t_{\rm age} \rm <6~Gyr$. Lastly, the galaxy m0204 has a lack of young-intermediate aged stars younger than $\rm 8~Gyr$. It does not show a distinct correlated AMR of young stars compared to the other galaxies.

 \begin{figure*}
 \plotone{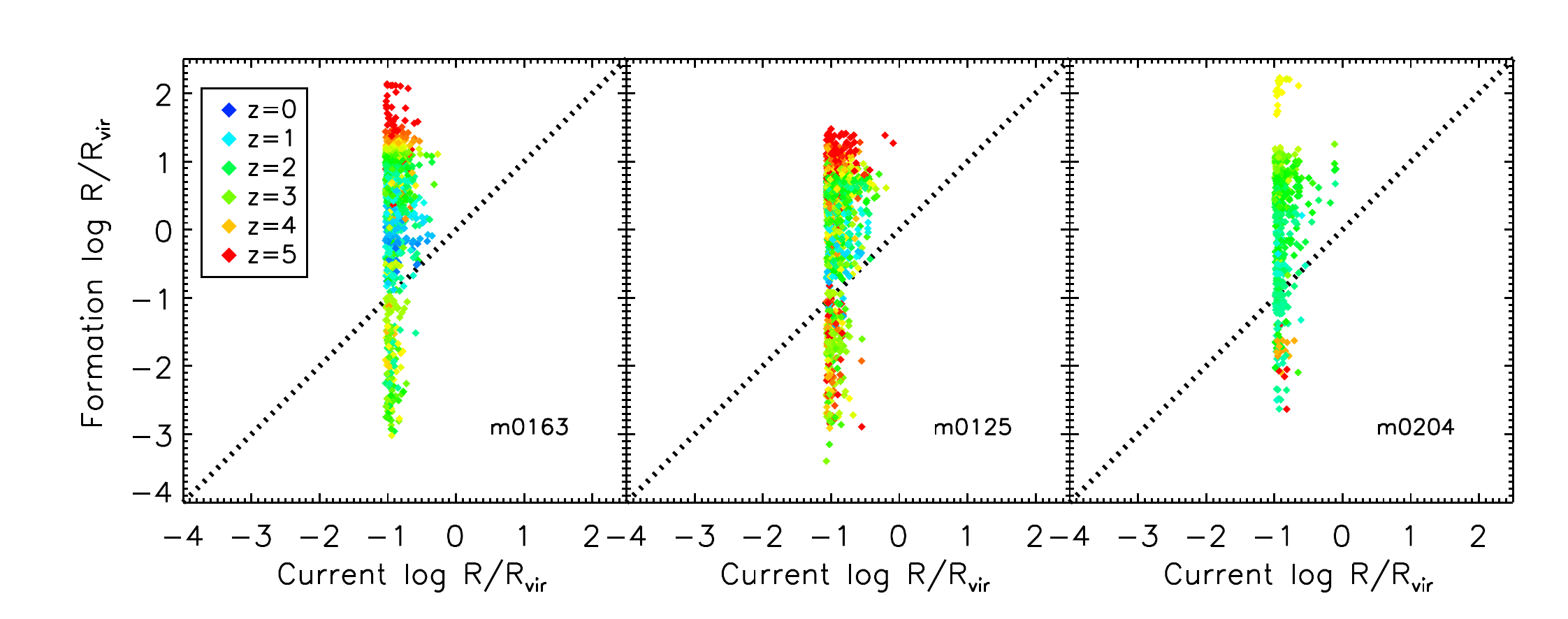}
 \caption{The radius where a star particle was born in units of the virial radius of the main galaxy at that time versus its current radius, for all stars within the outer halo field. The black dotted line indicates $x=y$, i.e., star particles below this line originally formed in the inner region but radially migrated  to the outer region. Color coding corresponds to the redshift when each star particle formed, ranging from $z=5$ (red/old) to $z=0$ (blue/young).\label{fig:radialmigration} }
\end{figure*}

Now we investigate the formation history of stars found in the outer halo. We follow back in time every star particle that ends up within a projected radius of $37.5 < r_{\rm proj} < 42.5$ kpc and $r<r_{\rm vir}$. We check both the formation time when a star particle is born and the formation location where it is born, i.e., its distance from the galaxy center. Figure~\ref{fig:origin} visualizes the stellar origin of our fiducial field at $R \sim 40$ kpc for three example galaxies. The color of each simulated star particle scales with its metallicity. The histograms for the formation radii for outer halo field stars show that massive simulated galaxies' outer halos consist mainly of accreted stars. 

If we take a conventional spatial division of 10\% of the virial radius $r_{10}$ for in situ vs. accreted \citep{Oser2010}, the fraction of mass in accreted stars is 87.35\%, 82.25\%, and 86.8\% respectively for the three galaxies. Most of these accreted stars are formed at relatively early times ($z>3$) in small systems outside the virial radius of the galaxy but added to the parent galaxy late in its evolution. As they originate from lower mass, lower metallicity systems, the bulk of them are metal poor. However, some of the accreted stars are younger and more metal-rich. The accreted stars formed as a consequence of the interactions with the main galaxy are clearly visible along the thin evolutionary tracks in Figure~\ref{fig:origin}. The younger stars are more metal-rich in these evolutionary tracks, which are most visible in the case of galaxy m0163. This indicates that these small systems experience a period of star formation and chemical enrichment as they merge into the main galaxy. Depending on its recent merger history, the age distribution of stars in outer fields differs for each example galaxy. For example, m0163, which had mergers most recently, shows more prominent young stellar populations.
 
 \begin{figure*}
 \plotone{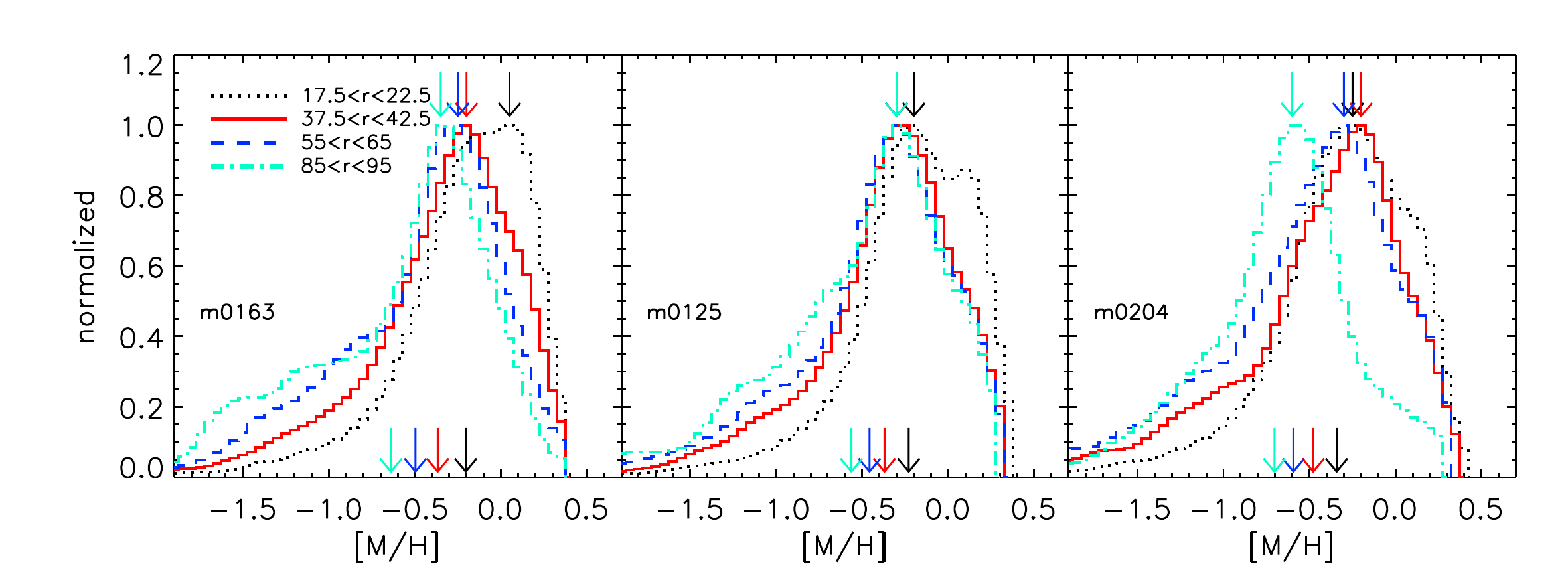}
 \caption{Normalized metallicity distribution functions of stars in projected fields centered at galactocentric radius of $20$ kpc (black dotted), $40$ kpc (red solid), $60$ kpc (blue dashed), and $90$ kpc (cyan dot-dashed) in example galaxies m0163, m0125, and m0204. The locations of the histogram peaks are marked with downward arrows on the top, and the median metallicity of stars in each field is shown with downward arrows near the $x$-axis. The simulated MDFs include a smoothing representing an estimated observational uncertainty of $\Delta {\rm [M/H]}=0.1$. \label{fig:radialgradient} }
\end{figure*}

\begin{figure*}
\plotone{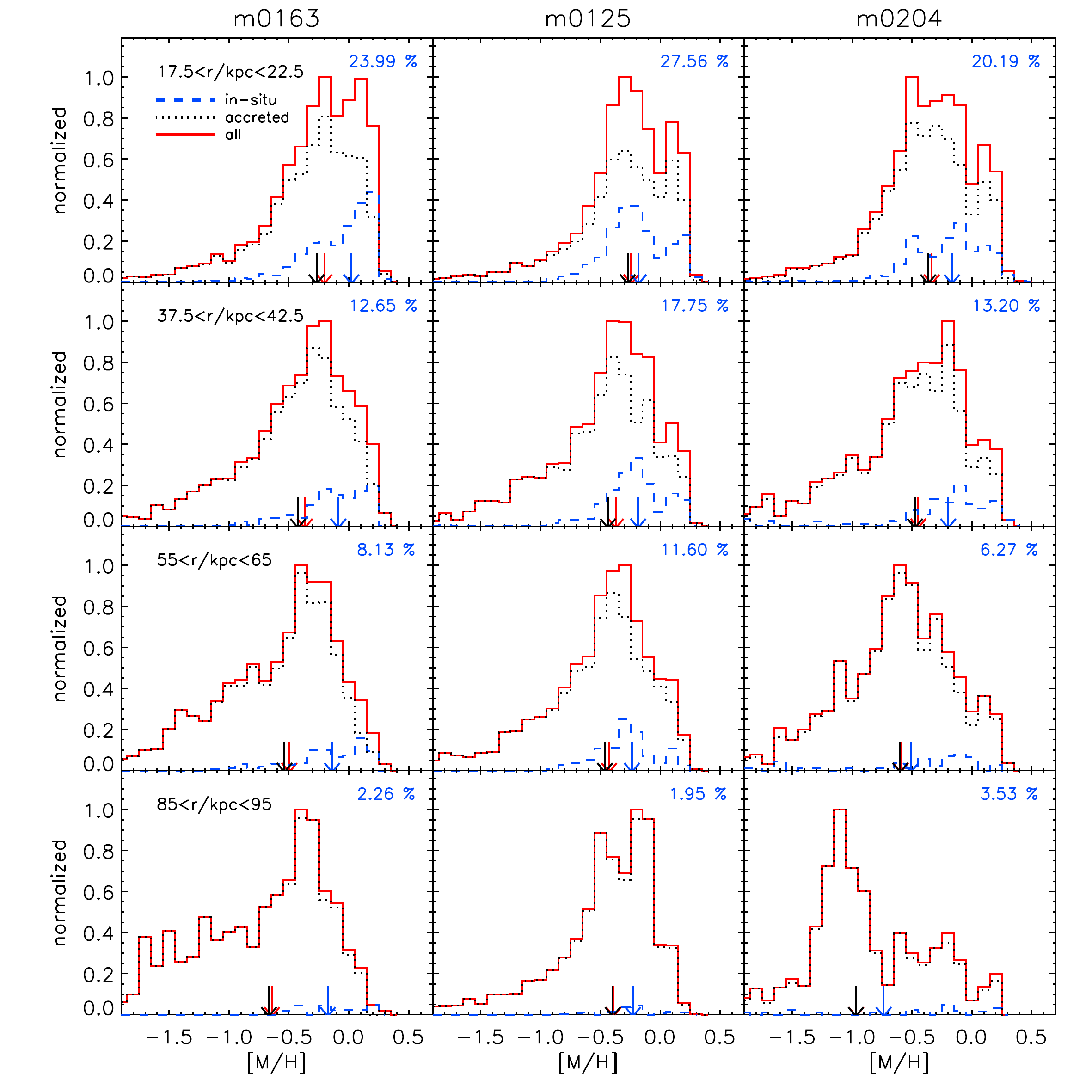}
\caption{MDF of all stars (solid red line) in projected fields centered at a galactocentric radius of $20$, $40$, $60$, and $90$ kpc in example galaxies m0163, m0125, and m0204. MDF of {\it in situ} formed stars (blue dashed) and MDF of those formed outside the galaxy and {\it accreted} later on (black dotted) are also shown. The downward arrows mark the median value of the stellar metallicity of each component. The fraction of in situ stars in each field is shown in each panel. The outer halo field of a massive elliptical galaxy mainly consists of accreted stars, and the contribution of in situ formed stars decreases with increasing galactocentric distance. The innermost halo region includes the largest fraction of in situ stars \label{fig:mdftwophaseradial} }
\end{figure*}

 \begin{figure*}
 \plotone{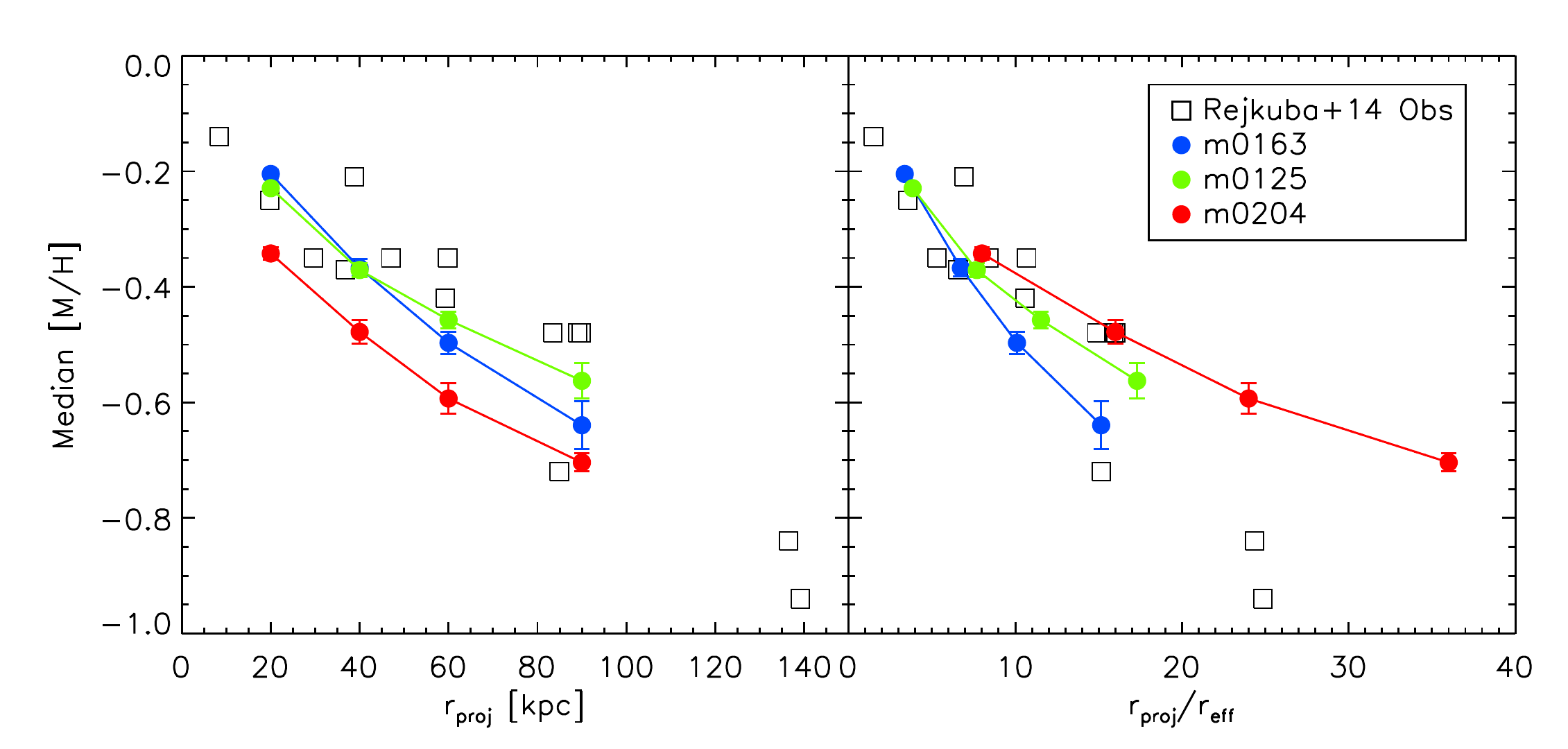}
 \caption{Median metallicity profile of stars as functions of projected distance in units of physical scale (left) and in units of the effective radius (right), for the simulated galaxies (filled circles). The uncertainties of the median metallicity are derived from bootstrap resampling and marked by error bars. Observational data (open squares) are the median metallicities of RGB stars in 14 different fields compiled by \cite{Rejkuba2014}. \label{fig:radialprofile} }
\end{figure*}

In order to show how the stars that populate the outer halo have migrated, we show the formation radii versus the final radii of stellar particles in the outer halo field of three example galaxies in Figure~\ref{fig:radialmigration}. As we select the outer halo stars within a projected distance of $\rm 37.5 < r_{\rm proj}/kpc < 42.5$, and within the virial radius $r<r_{\rm vir}$, the stars are distributed in a narrow range of final radii. However, as previously seen in Figure~\ref{fig:origin}, their formation radii are widely distributed. This is because the bulk of stars forms ex situ, outside of the galaxy at $r/r_{\rm vir} \sim 10^{0.5}$ to 10, and is accreted later on. Below the dotted line, in situ formed stars are shown, i.e., the stars that have formed within the galaxy ($r/r_{\rm vir} \sim 10^{-1}$) and then migrate outwards. They are currently located in a relatively inner region compared to ex situ formed stars. The presence in the outer halo of star particles that formed in the inner region can be attributed to radial heating and major mergers (see also the discussion in Sec.~\ref{sec:radialmigration}), but they mostly occupy the inner region of the selected outer halo fields.

So far, we have focused on the metallicity distribution of stars at $r_{\rm proj} \sim 40$ kpc, which corresponds to the target field of the \citet{Rejkuba2005} observations. Now we study the metallicity distribution of stars in fields at larger and smaller projected radii than $r_{\rm proj} \sim 40$ kpc. In Figure~\ref{fig:radialgradient}, we show normalized metallicity distribution functions of stars in four projected fields with average galactocentric radii of 20 kpc, 40 kpc, 60 kpc, and 90 kpc. Note that the MDFs include a smoothing of $\Delta {\rm [M/H]}=0.1$ to represent estimated observational uncertainties, and unsmoothed MDFs are shown in Figure~\ref{fig:mdftwophaseradial}. The median metallicity of stars decreases with increasing projected distance from the center of the galaxy. The peak or mode of the MDF also decreases with increasing projected radius. For example, the location of the peak shifts from $\rm [M/H]\sim0.02$ for the innermost region of $r_{\rm proj}=20$ kpc, to $\rm [M/H]\sim -0.38$ for the outermost region of $r_{\rm proj}=90$ kpc for the example galaxy m0163. The peak of the MDF is shifted even further for the galaxy m0204 to $\rm [M/H]\sim -0.6$. In the case of the galaxy m0125, the difference of the peak locations is subtle except for the innermost region of $r_{\rm proj}=20$ kpc. Metal poor tails dominate in the outer fields of all example galaxies.

In Figure~\ref{fig:mdftwophaseradial}, we separately show the MDF of accreted stars and that of in situ stars in each projected field at galactocentric radii 20 kpc, 40 kpc, 60 kpc, and 90 kpc for the three example galaxies. Compared to the main outer halo field of $37.5 < r_{\rm proj} < 42.5$ kpc (panels in the second row), the inner halo fields have more in situ formed stars. In the case of m0163, the in situ stars form a secondary metal-rich peak in the total MDF. The metal-rich secondary peaks in other galaxies are due to the shape of the MDF of the accreted component. As seen in Figure~\ref{fig:agehist}, some of the merged galaxies undergo fast chemical enrichment at early times, and add metal-rich stars to the final galaxy when they merge in. With increasing galactocentric distance, there is less contribution of in situ formed stars.

Figure~\ref{fig:radialprofile} shows the median metallicity as a function of projected distance in units of physical scale (left) and in units of the effective radius (right) for simulated galaxies and NGC 5128. We estimate the uncertainties of the median metallicity using bootstrap resampling \citep{efron1979}. The observational data is from \citet{Rejkuba2014}, who obtained the metallicity distributions of RGB stars out to a projected distance of 140 kpc, which corresponds to $25~r_{\rm eff}$ for NGC 5128. They also compiled the data from the previous NGC 5128 observations of \citet{Harris1998,Harris2000,Harris2002,Rejkuba2005} and showed the metallicity gradient with a more metal-rich inner region and lower metallicity outer halo, with some field-to-field metallicity variations (e.g. $r_{\rm proj} \sim 85$ kpc). We compare the median metallicity of stars in our simulated galaxies for four fields from 20 kpc to 90 kpc previously shown in Figure~\ref{fig:radialgradient} to the observed metallicity profile of \citet{Rejkuba2014} and find overall good agreement. As already shown in Figure~\ref{fig:mdfobs}, the galaxy m0204 has a lower median metallicity compared to NGC 5128, especially for the inner fields, for the same physical projected distance. However, the difference becomes negligible when comparing the fields' metallicity for the projected distance in units of effective radius. 

\section{Discussion}\label{sec:discussion}
\subsection{The stellar accretion origin of outer halo stars}
By tracing the origin of each stellar particle and checking its formation location, we have shown that $\sim 85$\% of the stars in the outer halo fields have an accreted origin (Figure~\ref{fig:mdftwophase} and \ref{fig:origin}). That means, most of the stars in the outer halos were not born there, but instead were born in the much lower mass systems that have been accreted and shredded in the stellar envelopes of the early-type galaxies as discussed in many previous studies \citep[e.g.][]{Naab2007,Bezanson2009,Oser2010,2013MNRAS.429.2924H,Hirschmann2013,VanDokkum2014,VanDokkum2015,Hirschmann2015}. The fraction of accreted stars is higher in the outer region than in the whole galaxy as dry minor mergers predominantly deposit stars in the outskirts of the galaxies (See Figure~\ref{fig:mdftwophaseradial}). These accreted stars are formed at relatively early times in lower mass, lower metallicity systems; therefore they are mostly old (see Figure~\ref{fig:agehist}) and metal poor. Especially, the metal-poor tail of the MDF in the outer halo is primarily populated by accreted stars in all three simulated galaxies (Figure~\ref{fig:mdftwophase}). This is consistent with the finding of \citet{Calura2012}, who showed that the low-metallicity tail (stars with $\rm [M/H]<-4$) of the simulated halo of a Milky Way mass galaxy mainly consists of accreted stars. However, other studies have shown that lower mass galaxies (such as the one studied by \citet{Calura2012}) have a more significant contribution from in situ star formation than the more massive gE type galaxies in our study \citep[see e.g.][]{Oser2010,Rodriguez-Gomez2016}.

All three galaxies did not have major mergers since $z=2$ but had many dry minor mergers of small systems, which mostly added low metallicity stars. We found that recent gas-rich mergers can deposit metal-rich and young stars in the outer halo region. m0163 had two recent gas-rich minor mergers at $z=0.2$ and $z=0.33$ that added newly formed stars. The merger at $z=0.33$ was more massive than another with mass ratio $(>1:8)$. m0175 also had a recent minor merger at $z=0.7$, adding some young stars formed during the interaction, and m0204 had no gas-rich mergers recently. Despite having different merger histories, all three galaxies show comparable MDFs. The age distribution of stars, however, is more sensitive to the recent merger histories (Figure~\ref{fig:agehist}) as found in \citet{DSouza2018a,DSouza2021}. We expect that the observed age distribution of outer fields can narrow the range of allowed interactions \citep[see also][]{Wang2020}.

\subsection{Contribution of in situ formed stars via radial migration}\label{sec:radialmigration}
The effect of radial migration of stars on the stellar metallicity distribution has been studied mainly for the Galactic disk, and it has been shown that stellar radial migration broadens MDFs and produces a greater dispersion \citep{DiMatteo2013,Grand2015,Halle2015a}. Here we focus on the effect of `outward' radial migration of centrally formed in situ stars.

Although the majority of stars in outer halos are accreted in our fiducial field at projected radius $R \sim 40$ kpc, $12\--17$\% of stars are formed in situ within the main progenitor in the simulations (Figure~\ref{fig:origin}). Some of them formed in the central region and then migrated out to larger radii (Figure~\ref{fig:radialmigration}). The majority of the centrally formed in situ stars remain in the central part of the galaxy \citep[see][]{Oser2010,Oser2011,Hirschmann2015}, but some of the in situ stars formed at high redshift can migrate over quite large distances, over 10 kpc since $z=2$ \citep[see their Figure 5]{Choi2018}. 

This migration of stars is a result of a combination of several physical processes. First of all, gas ejection by outflows can result in the expansion of the stars in the central region \citep{1980ApJ...235..986H}. Slow gas mass loss associated with the death of old stars such as AGB-winds can adiabatically expand the central region. Also, rapid gas mass loss via SN-driven and AGN-driven winds, which are very common in massive systems as shown in \citet{Brennan2018}, can cause an impulsive expansion \citep{1979ApJ...230L..33B,2008ApJ...689L.101F}. 

Second, the centrally formed stars can gain angular momentum via dynamical interactions during major mergers and migrate outward. For example, \citet{Hirschmann2015} showed that galaxies that have experienced recent major mergers can have a significant fraction of centrally formed in situ stars that have migrated outwards to the outskirts of galaxies. 

Lastly, a binary black hole formed by a merger of two galaxies scours the stars from the nucleus as the binary hardens \citep{2001ApJ...563...34M}. The ejection of surrounding stars enables the energy liberated by the shrinking of the binary to evacuate the central region and is believed to produce the observed cores in the light profiles of massive galaxies \citep[e.g.][]{1997AJ....114.1771F}. The simulations presented in this paper include black hole mergers and the associated heating effect of the shrinking orbit of a binary system. However, we expect the effect of the black hole binary scouring on the ejection of in situ formed stars to be negligible in the simulations due to the insufficient resolution.

Since recently formed in situ stars would not have had enough time to migrate out to larger radii, the in situ stars found in the outskirts of galaxies are primarily old (Figure~\ref{fig:agehist}). Even though they are born at high redshift around $z\sim3$, their metallicity is much higher than that of accreted stars (see Figure~\ref{fig:mdftwophase}, \ref{fig:origin}). This is because the central regions of the most massive progenitors underwent rapid chemical enrichment. In our fiducial outer field at $r_{\rm proj} \sim 40$~kpc of NGC 5128, the contribution of in situ stars that have migrated out to that distance is likely to be minor, but we expect that a metal-rich in situ star population can be dominant in the inner region of galaxy halos (Figure~\ref{fig:mdftwophaseradial}, and see more discussion in Section~\ref{sec:radial}).

\subsection{Variations in the MDF with radial distance} \label{sec:radial}
The study of metallicity gradients in elliptical galaxies has a long history since \citet{DeVaucouleurs1961}. Due to a decline in metallicity with increasing radius, elliptical galaxies are generally observed to be bluer at larger radii by photometric studies \citep{Strom1978,Zibetti2005,Suh2010,Tortora2010,Zibetti2020}. This is confirmed by spectroscopic studies that have looked at the gradients in the equivalent width of metal lines \citep{Fisher1995,Baes2007}. The metallicity gradients of elliptical galaxies have been typically studied within $r_{\rm eff}$, but recently observed out to larger radii $\sim 8 r_{\rm eff}$ \citep[e.g.][]{Coccato2010,LaBarbera2012,Greene2012,Greene2013}. Via resolved stellar photometry, \citet{Rejkuba2014} studied the metallicity gradient even further out, to $25 \thinspace r_{\rm eff}$. They obtained the metallicity distribution functions of individual RGB stars in each ACS field and investigated the halo metallicity gradient by measuring the median metallicity of stars as functions of the projected distance from the center of NGC 5128. We compare the observed median metallicity of stars from \citet{Rejkuba2014} to the simulations and find a consistent decline in metallicity with the projected distance (Figure~\ref{fig:radialprofile}). We find a change in the median [M/H], $\Delta \rm{[M/H]} \sim -0.3$, 0.2, and 0.23 dex over $40 - 90$ kpc for m0163, m0125 and m0204 respectively. This corresponds to a gradient of $\sim 0.006, 0.004, 0.0046$ dex $\rm {kpc^{-1}}$ which is comparable or modestly higher than the observed value found by \cite{Crnojevic2013} via a VLT/VIMOS survey.

Not only the median metallicity but the overall shape of MDFs differs with the projected distance. The MDF of the outermost halo field shows an extended metal-poor tail for all simulated galaxies (Figure~\ref{fig:mdftwophaseradial}). This is consistent with the finding of \cite{Harris2002}, who showed a longer and thicker tail toward lower metallicity in the MDF of the outer halo field compared to that of the inner halo field (see their Figure~9). \cite{Crnojevic2013} derived MDFs of NGC 5128 up to $\sim 85$ kpc, and also found a thicker metal-poor tail in the outer region. \cite{Lee2016} also showed that the outer halo field of a standard elliptical galaxy M105 exhibits an extended metal-poor tail compared to the inner region (see their Figure~12). 

In contrast, inner halos of elliptical galaxies are observed to have a prominent metal-rich population and sometimes to have two distinct sub-populations \citep{Harris2002,Lee2016}. Inner halos tend to have more contribution from centrally formed in situ stars via radial migration, sometimes showing a distinct metal-rich sub-population in MDFs (Figure~\ref{fig:radialgradient} and \ref{fig:mdftwophaseradial}). The population mixture of in situ and accreted stars can dominate the transition of MDF for the inner halo regions. This is consistent with the sharply declining radial number density profile of the metal-rich RGB stars in the outer region of M105 found in \cite{Lee2016}. \cite{Bird2014} found no evidence of a steep density falloff of metal-rich stars ([M/H]$\ge-0.7$) within 65 kpc in NGC 5128, but the transition must lie further out as in our simulated galaxies.

\subsection{Age distribution of outer halo stars}
Even though we applied a weighting function when we calculated the MDFs (see Section~\ref{subsec:method}), the stellar metallicity obtained in observations and that from simulations are derived in a fundamentally different fashion. In observations, photometric metallicities are derived by assuming a {\it single} age and a fixed $\alpha$-element abundance. Then individual metallicity values of stars are obtained by interpolating between model isochrones with different metallicities. For the simulations, we take the age of each star particle as shown in Figure~\ref{fig:agehist}. However, the discrepancy between observed and simulated age is expected to be minor, as the bulk of outer halo stars are born early with $t_{\rm age} \sim 10 $ Gyr in simulations, consistent with the commonly adopted stellar age of the isochrone fitting. 

We found that most stars are old in the outer halo field of simulated galaxies (Figure~\ref{fig:agehist}). \cite{Rejkuba2010} also found that the bulk of outer halo stars are old in the outer field of NGC  5128. They constructed synthetic color-magnitude diagrams (CMD) with varying star formation histories to find a matched CMD to the data and found that two burst models best fit the data. They found that $70-80$ \% of the stars are old with $t_{\rm age}\sim 12$ Gyr, and old stellar populations with the full metallicity range of the models ($Z=0.0001-0.04$) are needed to reproduce the observed RGB stars in their synthetic CMD models. This is consistent with our findings that all simulated galaxies have old stars that span a broad range in metallicity of $\rm -2.0 <[M/H]<0.3$ (Figure~\ref{fig:agehist}).

There are some contributions from young and intermediate-age stars in simulated galaxies. \cite{Rejkuba2010}  found that $20-30$ \% of stars are young with $t_{\rm age}\sim 2-4$ Gyr in the outer field of NGC 5128. \cite{Rejkuba2021}  recently confirmed this result, finding that a young stellar population with $t_{\rm age}\sim 4$ Gyr or younger is needed to reproduce the observed excess of stars above the tip of the RGB. Among our three simulated galaxies, only m0163 shows a considerable amount of young stars with $t_{\rm age} \sim 4 $ Gyr comparable to these recent findings.

\subsection{Caveats and limitations}
We archive excellent agreement between the simulation predictions for MDF and the observations of NGC 5128, but there remains uncertainty in the models. Adjusting the parameterization of the subgrid routines of stellar feedback and AGN feedback can alter the MDF outputs. AGN feedback is found to play the most dominant role in reproducing the fundamental physical properties of observed massive elliptical galaxies in the simulations \citep{Choi2017}. However, the stellar feedback is expected to play an essential role in reproducing MDF, as a metallicity of stars in low mass satellite galaxies that make up the ex situ stellar halo could be very sensitive to stellar feedback. Moreover, the physical mechanisms that are not included in the simulations, such as the effects of relativistic AGN jets or cosmic rays, can also affect the results. The significance of each effect on the resulting MDFs will be explored in detail in future work.

Although our simulations represent a good match to the observed MDF of NGC 5128, the current results may not be representative of massive elliptical galaxies in general. We present a minimal number of simulated samples applicable to a single observed galaxy. In order to draw general conclusions on massive elliptical galaxies and their MDFs, it is necessary to study a more significant number of simulated elliptical galaxy samples from a cosmological volume simulation such as Illustris-TNG \citep{nelson2018}, EAGLE \citep{schaye2015}, or SIMBA \citep{dave2019}. Analyzing simulated MDFs of elliptical galaxies with different star formation and stellar assembly histories and group environments will help us better understand the complex formation history of elliptical galaxies. We reserve this for future work.

As discussed in Sec.~\ref{subsec:method}, we did not use a small target field to select the outer halo stars as in observations due to the limited resolution. Instead, we use all stellar particles within a projected annulus to sample enough star particles to obtain MDFs. Although sub-structure is expected to be present in the outer halo, and \cite{Rejkuba2014} also showed the large field-by-field variation in the outer field of galaxies, field-to-field scatter cannot be studied with current simulations due to small number statistics. To check the variations between different fields in the outer halo, much higher resolution simulations will be needed, which is beyond the scope of this paper. This subject is reserved for future work.

\subsection{Future perspective}
A detailed study of the stellar populations in nearby elliptical galaxies will allow us to understand better the physical processes involved in the formation and growth of massive elliptical galaxies. The analysis of the color-magnitude diagram of the resolved stars is the most direct way to estimate the metallicity and age of stellar populations. However, the amount of data is still limited, as it has to come from HST or a ground-based telescope under very good seeing conditions \citep{Crnojevic2013}. Therefore, most of the stellar mass of NGC 5128 is still left unexplored. With the James Webb Space Telescope (JWST) and the next generation of large-aperture ground-based telescopes, it will be possible to map the metallicity distribution over the whole galaxy, even for the crowded inner regions \citep{Schreiber2014}. 

JWST's ability to resolve individual stars will offer the opportunity to perform accurate stellar photometry in crowded fields of local galaxies which are within reach of the HST. Moreover, with its much higher spatial resolution and higher sensitivity, JWST will bring more distant targets within reach, allowing us to study in detail the stellar populations in distant ellipticals to the Leo group objects at $D \sim 10$ Mpc, and then outward to Virgo ($D \sim 16$ Mpc) and Fornax ($D \sim 19$ Mpc). By resolving individual RGB stars and studying their calibrated infrared features, one can determine their metallicity distribution \citep[e.g.][]{Valenti2006}. As the ellipticals may result from a wide range of formation histories, a detailed comparison of metallicity distributions for a more significant number of observed elliptical samples to models will allow us to better understand  how ellipticals formed and have evolved.

\section{Summary}
We investigate the metallicity distribution functions of stars in the halos of simulated elliptical galaxies and compare them to the resolved stellar photometry observations of the local gE galaxy NGC 5128. Our main findings are summarized as follows:

\begin{itemize}
\item The predicted MDFs from simulations are in good qualitative agreement with the observed MDF of NGC 5128. The correspondence between simulated and observed MDFs indicates that the treatment of metal production, expulsion, and accretion of simulations is approximately correct. 

\item Using cosmological simulations of massive galaxy formation, we predict that most of the stars in outer fields of massive galaxies have an `accreted' origin, formed in other small galaxies and later merged in. 
\item The contribution of centrally formed in situ stars is sub-dominant in the outer halo but can be dominant in the inner regions as they can more effectively migrate outward from the center.
\item We find a decline in median metallicity with the projected galactocentric distance, consistent with the observed median metallicity of stars from \citet{Rejkuba2014}. We also find that the overall shape of MDFs differs with the projected distance, i.e., the MDF of the outer halo field shows a more extended metal-poor tail.
\item This study presents a limited number of simulated galaxy samples compared to a single observed galaxy NGC 5128. However, the results presented here can provide the baseline for future studies on comparing MDFs in simulations and observations of elliptical galaxies. Resolved stellar photometry from upcoming telescope facilities such as JWST will provide more constraints for galaxy formation models. Combining these data with more galaxy samples from larger volume simulations will enable us to refine our models of massive galaxy formation.
\end{itemize}

\begin{acknowledgments}
The authors would like to thank the anonymous referee for constructive suggestions and comments. We are grateful to Charlie Conroy and Jaehyun Lee for their helpful discussions. We also thank Marina Rejkuba for kindly providing the table of observational data. Numerical simulations were run on the computer clusters of the Princeton Institute of Computational Science and engineering. The authors acknowledge the Office of Advanced Research Computing (OARC) at Rutgers, The State University of New Jersey, for providing access to the Amarel cluster and associated research computing resources that have contributed to the results reported here (URL: http://oarc.rutgers.edu). RSS acknowledges support from the Simons Foundation. MH acknowledges financial support from the Carlsberg Foundation via a ``Semper Ardens'' grant (CF15-0384).
\end{acknowledgments}

\bibliography{library}

\end{document}

%% file: defs.tex
\def\galaxiespprox{\mathrel{\vcenter{\offinterlineskip \hbox{$>$}
    \kern 0.3ex \hbox{$\sim$}}}}
\def\lapprox{\mathrel{\vcenter{\offinterlineskip \hbox{$<$}
    \kern 0.3ex \hbox{$\sim$}}}}
\newcommand{\beq}{\begin{equation}} 
\newcommand{\eeq}{\end{equation}}

\def\eps{\epsilon}

\def\Sig15{\Sigma_{1.5}}

\def\kpc{{\rm\thinspace kpc}}

\def\Msun{\hbox{$\thinspace \rm M_{\odot}$}}
\def\Msunh{\hbox{$\thinspace  \it h^{\rm -1} \rm \thinspace M_{\odot} $}}

\def\strom{Str$\rm \ddot{o}$mgren$\thinspace$}

\def\delm12{M_{12}}
\def\sigm1{\sigma(m_{1})}

%% file: tab1.tex
\begin{table}
   \begin{center}
   \caption{Overview of physical quantities of NGC 5128 and three simulated galaxies. \label{tab:summary}}
   \vskip+0.5truecm
    {
   \begin{tabular}{c|c|ccc}\hline
 &  Observation & \multicolumn{3}{c}{Simulation}\\
   
   &
NGC 5128 & 
m0163 & 
m0125 & 
m0204 \cr\hline
$M_{\rm vir}$ ($M_{\odot}$) & $10^{12.96^{+0.12}_{-0.17}}$  $^{a}$ & $10^{12.9}$ & $10^{13.0}$ & $10^{12.8}$\cr
$M_{\rm star}$ ($M_{\odot}$) & $10^{11.46 \pm 0.12}$ $^{b}$ & $10^ {11.5}$& $10^{11.7}$ & $10^{11.6}$ \cr
$r_{\rm vir}$ (kpc) & $400 \pm 43$$^{a}$ & 407.6& 441.7 & 368.8 \cr
$r_{\rm eff}$ (kpc) & $5.6 \pm 0.15$$^{c}$ & 5.94 & 5.20& 2.50 \cr
  \hline\hline
   \end{tabular}}
   \end{center}
      \tablecomments{
    $^{a}$ This is the total mass of Cen A group, measured with dynamics of satellites \citep{Woodley2006}. The Cen A group virial radius is estimated accordingly ($r_{\rm vir} \equiv r_{200}$) \citep{Rejkuba2014}.\\ 
    $^{b}$  We adopted a distance modulus $(m-M)_V=27.91$ from  \citep{Harris2009}, and estimated the stellar mass using an integrated magnitude of $V^T=6.2$ \citep{VandenBergh1976}, and $(M/L)_V = 7$ following \citet{Harris2002}.\\
    $^{c}$ The half light radius of NGC 5128 is $R_{\rm eff}=305''$ which corresponds to 5.6 kpc at the 3.8 Mpc distance \citep{Harris2009}.\\    
   }
\end{table}

%% file: ms_gdwarf_MDF_arxiv.bbl
\begin{thebibliography}{}
\expandafter\ifx\csname natexlab\endcsname\relax\def\natexlab#1{#1}\fi

\bibitem[{Anders {et~al.}(2014)Anders, Chiappini, Santiago, Rocha-Pinto,
  Girardi, {Da Costa}, Maia, Steinmetz, Minchev, Schultheis, Boeche, Miglio,
  Montalb{\'{a}}n, Schneider, Beers, Cunha, {Allende Prieto}, Balbinot,
  Bizyaev, Brauer, Brinkmann, Frinchaboy, {Garc{\'{i}}a P{\'{e}}rez}, Hayden,
  Hearty, Holtzman, Johnson, Kinemuchi, Majewski, Malanushenko, Malanushenko,
  Nidever, O'Connell, Pan, Robin, Schiavon, Shetrone, Skrutskie, Smith,
  Stassun, \& Zasowski}]{Anders2014a}
Anders, F., Chiappini, C., Santiago, B.~X., {et~al.} 2014, Astronomy and
  Astrophysics, 564, 28

\bibitem[{Arav {et~al.}(2020)Arav, Xu, Miller, Kriss, \& Plesha}]{Arav2020}
Arav, N., Xu, X., Miller, T., Kriss, G.~A., \& Plesha, R. 2020, The
  Astrophysical Journal Supplement Series, 247, 37

\bibitem[{Asplund {et~al.}(2009)Asplund, Grevesse, Sauval, \&
  Scott}]{Asplund2009}
Asplund, M., Grevesse, N., Sauval, A.~J., \& Scott, P. 2009, Annual Review of
  Astronomy and Astrophysics, 47, 481

\bibitem[{Aumer {et~al.}(2013)Aumer, White, Naab, \&
  Scannapieco}]{2013MNRAS.434.3142A}
Aumer, M., White, S.~D., Naab, T., \& Scannapieco, C. 2013, Monthly Notices of
  the Royal Astronomical Society, 434, 3142

\bibitem[{Baes {et~al.}(2007)Baes, Sil'chenko, Moiseev, \& Manakova}]{Baes2007}
Baes, M., Sil'chenko, O.~K., Moiseev, A.~V., \& Manakova, E.~A. 2007, Astronomy
  and Astrophysics, 467, 991

\bibitem[{Bekki {et~al.}(2002)Bekki, Harris, \& Harris}]{Bekki2002}
Bekki, K., Harris, W.~E., \& Harris, G. L.~H. 2002, Monthly Notices of the
  Royal Astronomical Society, 338, 587

\bibitem[{Bezanson {et~al.}(2009)Bezanson, van Dokkum, Tal, Marchesini, Kriek,
  Franx, \& Coppi}]{Bezanson2009}
Bezanson, R., van Dokkum, P.~G., Tal, T., {et~al.} 2009, The Astrophysical
  Journal, 697, 1290

\bibitem[{Biermann \& Shapiro(1979)}]{1979ApJ...230L..33B}
Biermann, P., \& Shapiro, S. 1979, The Astrophysical Journal Letters, 230, L33

\bibitem[{Bird {et~al.}(2015)Bird, Flynn, Harris, \& Valtonen}]{Bird2014}
Bird, S.~A., Flynn, C., Harris, W.~E., \& Valtonen, M. 2015, Astronomy and
  Astrophysics, 575, 72

\bibitem[{Bondi(1952)}]{1952MNRAS.112..195B}
Bondi, H. 1952, Monthly Notices of the Royal Astronomical Society, 112, 195

\bibitem[{Bondi \& Hoyle(1944)}]{1944MNRAS.104..273B}
Bondi, H., \& Hoyle, F. 1944, Monthly Notices of the Royal Astronomical
  Society, 104, 273

\bibitem[{Brennan {et~al.}(2018)Brennan, Choi, Somerville, Hirschmann, Naab, \&
  Ostriker}]{Brennan2018}
Brennan, R., Choi, E., Somerville, R.~S., {et~al.} 2018, The Astrophysical
  Journal, 860, 14

\bibitem[{Calura \& Menci(2009)}]{Calura2009}
Calura, F., \& Menci, N. 2009, Monthly Notices of the Royal Astronomical
  Society, 400, 1347

\bibitem[{Calura {et~al.}(2012)Calura, Gibson, Michel-Dansac, Stinson, Cignoni,
  Dotter, Pilkington, House, Brook, Few, Bailin, Couchman, \&
  Wadsley}]{Calura2012}
Calura, F., Gibson, B.~K., Michel-Dansac, L., {et~al.} 2012, Monthly Notices of
  the Royal Astronomical Society, 427, 1401

\bibitem[{Carrera {et~al.}(2008)Carrera, Gallart, Hardy, Aparicio, \&
  Zinn}]{Carrera2008}
Carrera, R., Gallart, C., Hardy, E., Aparicio, A., \& Zinn, R. 2008, The
  Astronomical Journal, 135, 836

\bibitem[{Choi {et~al.}(2020)Choi, Brennan, Somerville, Ostriker, Hirschmann,
  \& Naab}]{Choi2020}
Choi, E., Brennan, R., Somerville, R.~S., {et~al.} 2020, The Astrophysical
  Journal, 904, 8

\bibitem[{Choi {et~al.}(2012)Choi, Ostriker, Naab, \& Johansson}]{Choi2012a}
Choi, E., Ostriker, J.~P., Naab, T., \& Johansson, P.~H. 2012, Astrophysical
  Journal, 754, 125

\bibitem[{Choi {et~al.}(2015)Choi, Ostriker, Naab, Oser, \& Moster}]{Choi2015a}
Choi, E., Ostriker, J.~P., Naab, T., Oser, L., \& Moster, B.~P. 2015, Monthly
  Notices of the Royal Astronomical Society, 449, 4105

\bibitem[{Choi {et~al.}(2017)Choi, Ostriker, Naab, Somerville, Hirschmann,
  N{\'{u}}{\~{n}}ez, Hu, \& Oser}]{Choi2017}
Choi, E., Ostriker, J.~P., Naab, T., {et~al.} 2017, The Astrophysical Journal,
  844, 31

\bibitem[{Choi {et~al.}(2018)Choi, Somerville, Ostriker, Naab, \&
  Hirschmann}]{Choi2018}
Choi, E., Somerville, R.~S., Ostriker, J.~P., Naab, T., \& Hirschmann, M. 2018,
  The Astrophysical Journal, 866, 91

\bibitem[{Coccato {et~al.}(2010)Coccato, Gerhard, \& Arnaboldi}]{Coccato2010}
Coccato, L., Gerhard, O., \& Arnaboldi, M. 2010, Monthly Notices of the Royal
  Astronomical Society, 407, L26

\bibitem[{Cohen {et~al.}(2020)Cohen, Goudfrooij, Correnti, Gnedin, Harris,
  Chandar, Puzia, \& S{\'{a}}nchez-Janssen}]{Cohen2020}
Cohen, R.~E., Goudfrooij, P., Correnti, M., {et~al.} 2020, The Astrophysical
  Journal, 890, 52

\bibitem[{Crnojevi{\'{c}} {et~al.}(2013)Crnojevi{\'{c}}, Ferguson, Irwin,
  Bernard, Arimoto, Jablonka, \& Kobayashi}]{Crnojevic2013}
Crnojevi{\'{c}}, D., Ferguson, A. M.~N., Irwin, M.~J., {et~al.} 2013, Monthly
  Notices of the Royal Astronomical Society, 432, 832

\bibitem[{Cullen \& Dehnen(2010)}]{2010MNRAS.408..669C}
Cullen, L., \& Dehnen, W. 2010, Monthly Notices of the Royal Astronomical
  Society, 408, 669

\bibitem[{Dav{\'{e}} {et~al.}(2019)Dav{\'{e}}, Angl{\'{e}}s-Alc{\'{a}}zar,
  Narayanan, Li, Rafieferantsoa, \& Appleby}]{dave2019}
Dav{\'{e}}, R., Angl{\'{e}}s-Alc{\'{a}}zar, D., Narayanan, D., {et~al.} 2019,
  Monthly Notices of the Royal Astronomical Society, 486, 2827

\bibitem[{de~Vaucouleurs(1961)}]{DeVaucouleurs1961}
de~Vaucouleurs, G. 1961, The Astrophysical Journal Supplement, 5, 233

\bibitem[{Dekel {et~al.}(2009)Dekel, Birnboim, Engel, Freundlich, Goerdt,
  Mumcuoglu, Neistein, Pichon, Teyssier, Zinger, Dekel, Birnboim, Engel,
  Freundlich, Goerdt, Mumcuoglu, Neistein, Pichon, Teyssier, \&
  Zinger}]{Dekel2009}
Dekel, A., Birnboim, Y., Engel, G., {et~al.} 2009, Nature, 457, 451

\bibitem[{Demarque {et~al.}(2004)Demarque, Woo, Kim, \& Yi}]{Demarque2004}
Demarque, P., Woo, J., Kim, Y., \& Yi, S.~K. 2004, The Astrophysical Journal
  Supplement Series, 155, 667

\bibitem[{{Di Matteo} {et~al.}(2013){Di Matteo}, Haywood, Combes, Semelin, \&
  Snaith}]{DiMatteo2013}
{Di Matteo}, P., Haywood, M., Combes, F., Semelin, B., \& Snaith, O.~N. 2013,
  Astronomy {\&} Astrophysics, 553, A102

\bibitem[{D'Souza \& Bell(2018)}]{DSouza2018a}
D'Souza, R., \& Bell, E.~F. 2018, NatAs, 2, 737

\bibitem[{D'Souza \& Bell(2021)}]{DSouza2021}
---. 2021, Monthly Notices of the Royal Astronomical Society, 504, 5270

\bibitem[{Durrell {et~al.}(2010)Durrell, Sarajedini, \& Chandar}]{Durrell2010}
Durrell, P.~R., Sarajedini, A., \& Chandar, R. 2010, The Astrophysical Journal,
  718, 1118

\bibitem[{Efron(1979)}]{efron1979}
Efron, B. 1979, The Annals of Statistics, 7, 1

\bibitem[{Eggen {et~al.}(1962)Eggen, Lynden-Bell, Sandage, Eggen, Lynden-Bell,
  \& Sandage}]{Eggen1962}
Eggen, O.~J., Lynden-Bell, D., Sandage, A.~R., {et~al.} 1962, The Astrophysical
  Journal, 136, 748

\bibitem[{Escala {et~al.}(2018)Escala, Wetzel, Kirby, Hopkins, Ma, Wheeler,
  Kere{\v{s}}, Faucher-Gigu{\`{e}}re, \& Quataert}]{Escala2018}
Escala, I., Wetzel, A., Kirby, E.~N., {et~al.} 2018, Monthly Notices of the
  Royal Astronomical Society, 474, 2194

\bibitem[{Faber {et~al.}(1997)Faber, Tremaine, Ajhar, Byun, Dressler, Gebhardt,
  Grillmair, Kormendy, Lauer, \& Richstone}]{1997AJ....114.1771F}
Faber, S., Tremaine, S., Ajhar, E., {et~al.} 1997, The Astronomical Journal,
  114, 1771

\bibitem[{Fan {et~al.}(2008)Fan, Lapi, {De Zotti}, \&
  Danese}]{2008ApJ...689L.101F}
Fan, L., Lapi, A., {De Zotti}, G., \& Danese, L. 2008, The Astrophysical
  Journal Letters, 689, L101

\bibitem[{Fisher {et~al.}(1995)Fisher, Franx, \& Illingworth}]{Fisher1995}
Fisher, D., Franx, M., \& Illingworth, G. 1995, The Astrophysical Journal, 448,
  119

\bibitem[{Grand {et~al.}(2015)Grand, Kawata, \& Cropper}]{Grand2015}
Grand, R.~J., Kawata, D., \& Cropper, M. 2015, Monthly Notices of the Royal
  Astronomical Society, 447, 4018

\bibitem[{Greene {et~al.}(2012)Greene, Murphy, Comerford, Gebhardt, \&
  Adams}]{Greene2012}
Greene, J.~E., Murphy, J.~D., Comerford, J.~M., Gebhardt, K., \& Adams, J.~J.
  2012, Astrophysical Journal, 750, 32

\bibitem[{Greene {et~al.}(2013)Greene, Murphy, Graves, Gunn, Raskutti,
  Comerford, \& Gebhardt}]{Greene2013}
Greene, J.~E., Murphy, J.~D., Graves, G.~J., {et~al.} 2013, The Astrophysical
  Journal, 776, 64

\bibitem[{Halle {et~al.}(2015)Halle, {Di Matteo}, Haywood, \&
  Combes}]{Halle2015a}
Halle, A., {Di Matteo}, P., Haywood, M., \& Combes, F. 2015, Astronomy and
  Astrophysics, 578, 58

\bibitem[{Harris \& Harris(2000)}]{Harris2000}
Harris, G. L.~H., \& Harris, W.~E. 2000, The Astronomical Journal, 120, 2423

\bibitem[{Harris {et~al.}(1999)Harris, Harris, \& Poole}]{Harris1998}
Harris, G. L.~H., Harris, W.~E., \& Poole, G.~B. 1999, The Astronomical
  Journal, 117, 855

\bibitem[{Harris {et~al.}(2010)Harris, Rejkuba, \& Harris}]{Harris2009}
Harris, G. L.~H., Rejkuba, M., \& Harris, W.~E. 2010, Publications of the
  Astronomical Society of Australia, 27, 457

\bibitem[{Harris \& Harris(2002)}]{Harris2002}
Harris, W.~E., \& Harris, G. L.~H. 2002, The Astronomical Journal, 123, 3108

\bibitem[{Harris {et~al.}(2007{\natexlab{a}})Harris, Harris, Layden, \&
  Stetson}]{Harris2007}
Harris, W.~E., Harris, G. L.~H., Layden, A.~C., \& Stetson, P.~B.
  2007{\natexlab{a}}, The Astronomical Journal, 134, 43

\bibitem[{Harris {et~al.}(2007{\natexlab{b}})Harris, Harris, Layden, \&
  Wehner}]{Harris2007a}
Harris, W.~E., Harris, G. L.~H., Layden, A.~C., \& Wehner, E.~H.
  2007{\natexlab{b}}, The Astrophysical Journal, 666, 903

\bibitem[{Hayden {et~al.}(2015)Hayden, Bovy, Holtzman, Nidever, Bird, Weinberg,
  Andrews, Majewski, Prieto, Anders, Beers, Bizyaev, Chiappini, Cunha,
  Frinchaboy, Garc{\'{i}}a-Her{\'{n}}andez, {Garc{\'{i}}a P{\'{e}} Rez},
  Girardi, Harding, Hearty, Johnson, M{\'{e}}sz{\'{a}}ros, Minchev, O'Connell,
  Pan, Robin, Schiavon, Schneider, Schultheis, Shetrone, Skrutskie, Steinmetz,
  Smith, Wilson, Zamora, \& Zasowski}]{Hayden2015}
Hayden, M.~R., Bovy, J., Holtzman, J.~A., {et~al.} 2015, Astrophysical Journal,
  808, 132

\bibitem[{Helmi {et~al.}(2006)Helmi, Irwin, Tolstoy, Battaglia, Hill, Jablonka,
  Venn, Shetrone, Letarte, Arimoto, Abel, Francois, Kaufer, Primas, Sadakane,
  \& Szeifert}]{Helmi2006}
Helmi, A., Irwin, M.~J., Tolstoy, E., {et~al.} 2006, The Astrophysical Journal,
  651, L121

\bibitem[{Hill {et~al.}(2011)Hill, Lecureur, Gomez, Zoccali, Schultheis,
  Babusiaux, Royer, Barbuy, Arenou, Minniti, \& Ortolani}]{Hill2011}
Hill, V., Lecureur, A., Gomez, A., {et~al.} 2011, Astronomy and Astrophysics,
  534, 80

\bibitem[{Hills(1980)}]{1980ApJ...235..986H}
Hills, J. 1980, The Astrophysical Journal, 235, 986

\bibitem[{Hilz {et~al.}(2013)Hilz, Naab, \& Ostriker}]{2013MNRAS.429.2924H}
Hilz, M., Naab, T., \& Ostriker, J. 2013, Monthly Notices of the Royal
  Astronomical Society, 429, 2924

\bibitem[{Hirschmann {et~al.}(2013)Hirschmann, Naab, Dav{\'{e}}, Oppenheimer,
  Ostriker, Somerville, Oser, Genzel, Tacconi, F{\"{o}}rster-Schreiber,
  Burkert, \& Genel}]{Hirschmann2013}
Hirschmann, M., Naab, T., Dav{\'{e}}, R., {et~al.} 2013, Monthly Notices of the
  Royal Astronomical Society, 436, 2929

\bibitem[{Hirschmann {et~al.}(2015)Hirschmann, Naab, Ostriker, Forbes, Duc,
  Dav{\'{e}}, Oser, Karabal, Hirschmann, Naab, Ostriker, Forbes, Duc,
  Dav{\'{e}}, Oser, \& Karabal}]{Hirschmann2015}
Hirschmann, M., Naab, T., Ostriker, J.~P., {et~al.} 2015, MNRAS, 449, 528

\bibitem[{Homma {et~al.}(2015)Homma, Murayama, Kobayashi, \&
  Taniguchi}]{Homma2015}
Homma, H., Murayama, T., Kobayashi, M.~A., \& Taniguchi, Y. 2015, Astrophysical
  Journal, 799, 230

\bibitem[{Hoyle \& Lyttleton(1939)}]{1939PCPS...34..405H}
Hoyle, F., \& Lyttleton, R. 1939, Proceedings of the Cambridge Philosophical
  Society, 34, 405

\bibitem[{Hu {et~al.}(2014)Hu, Naab, Walch, Moster, \&
  Oser}]{2014MNRAS.443.1173H}
Hu, C.~Y., Naab, T., Walch, S., Moster, B.~P., \& Oser, L. 2014, Monthly
  Notices of the Royal Astronomical Society, 443, 1173

\bibitem[{Hui {et~al.}(1993)Hui, Ford, Ciardullo, \& Jacoby}]{Hui1993}
Hui, X., Ford, H.~C., Ciardullo, R., \& Jacoby, G.~H. 1993, The Astrophysical
  Journal, 414, 463

\bibitem[{Israel(1998)}]{Israel1998}
Israel, F.~P. 1998, Astronomy and Astrophysics Review, 8, 237

\bibitem[{Iwamoto {et~al.}(1999)Iwamoto, Brachwitz, Nomoto, Kishimoto, Umeda,
  Hix, \& Thielemann}]{1999ApJS..125..439I}
Iwamoto, K., Brachwitz, F., Nomoto, K., {et~al.} 1999, The Astrophysical
  Journal Supplement Series, 125, 439

\bibitem[{Jeon {et~al.}(2017)Jeon, Besla, \& Bromm}]{Jeon2017}
Jeon, M., Besla, G., \& Bromm, V. 2017, The Astrophysical Journal, 848, 85

\bibitem[{Karakas(2010)}]{2010MNRAS.403.1413K}
Karakas, A.~I. 2010, Monthly Notices of the Royal Astronomical Society, 403,
  1413

\bibitem[{Kroupa(2001)}]{2001MNRAS.322..231K}
Kroupa, P. 2001, Monthly Notices of the Royal Astronomical Society, 322, 231

\bibitem[{{La Barbera} {et~al.}(2012){La Barbera}, Ferreras, de~Carvalho,
  Bruzual, Charlot, Pasquali, \& Merlin}]{LaBarbera2012}
{La Barbera}, F., Ferreras, I., de~Carvalho, R.~R., {et~al.} 2012, Monthly
  Notices of the Royal Astronomical Society, 426, 2300

\bibitem[{Larson(1975)}]{Larson1975}
Larson, R.~B. 1975, Monthly Notices of the Royal Astronomical Society, 173, 671

\bibitem[{Lee \& Jang(2016)}]{Lee2016}
Lee, M.~G., \& Jang, I.~S. 2016, The Astrophysical Journal, 822, 70

\bibitem[{Mackereth {et~al.}(2019)Mackereth, Schiavon, Pfeffer, Hayes, Bovy,
  Anguiano, Prieto, Hasselquist, Holtzman, Johnson, Majewski, O'connell,
  Shetrone, Tissera, \& Fer{\'{n}}andez-Trincado}]{Mackereth2019}
Mackereth, J.~T., Schiavon, R.~P., Pfeffer, J., {et~al.} 2019, Monthly Notices
  of the Royal Astronomical Society, 482, 3426

\bibitem[{Milosavljevi{\'{c}} \& Merritt(2001)}]{2001ApJ...563...34M}
Milosavljevi{\'{c}}, M., \& Merritt, D. 2001, The Astrophysical Journal, 563,
  34

\bibitem[{Monachesi {et~al.}(2016)Monachesi, Bell, Radburn-Smith, Bailin,
  de~Jong, Holwerda, Streich, \& Silverstein}]{Monachesi2016}
Monachesi, A., Bell, E.~F., Radburn-Smith, D.~J., {et~al.} 2016, Monthly
  Notices of the Royal Astronomical Society, 457, 1419

\bibitem[{Mould \& Spitler(2010)}]{Mould2010}
Mould, J., \& Spitler, L. 2010, Astrophysical Journal, 722, 721

\bibitem[{Naab {et~al.}(2007)Naab, Johansson, Ostriker, Efstathiou, Naab,
  Johansson, Ostriker, \& Efstathiou}]{Naab2007}
Naab, T., Johansson, P.~H., Ostriker, J.~P., {et~al.} 2007, ApJ, 658, 710

\bibitem[{Naab \& Ostriker(2017)}]{2017ARA&amp;A..55...59N}
Naab, T., \& Ostriker, J.~P. 2017, Annual Review of Astronomy and Astrophysics,
  55, 59

\bibitem[{Nelson {et~al.}(2018)Nelson, Pillepich, Springel, Weinberger,
  Hernquist, Pakmor, Genel, Torrey, Vogelsberger, Kauffmann, Marinacci, \&
  Naiman}]{nelson2018}
Nelson, D., Pillepich, A., Springel, V., {et~al.} 2018, Monthly Notices of the
  Royal Astronomical Society, 475, 624

\bibitem[{N{\'{u}}{\~{n}}ez {et~al.}(2017)N{\'{u}}{\~{n}}ez, Ostriker, Naab,
  Oser, Hu, \& Choi}]{2017ApJ...836..204N}
N{\'{u}}{\~{n}}ez, A., Ostriker, J.~P., Naab, T., {et~al.} 2017, The
  Astrophysical Journal, 836, 204

\bibitem[{Oser {et~al.}(2012)Oser, Naab, Ostriker, \& Johansson}]{Oser2011}
Oser, L., Naab, T., Ostriker, J.~P., \& Johansson, P.~H. 2012, The
  Astrophysical Journal, 744, 63

\bibitem[{Oser {et~al.}(2010)Oser, Ostriker, Naab, Johansson, \&
  Burkert}]{Oser2010}
Oser, L., Ostriker, J.~P., Naab, T., Johansson, P.~H., \& Burkert, A. 2010,
  Astrophysical Journal, 725, 2312

\bibitem[{Pagel(1989)}]{Pagel1989}
Pagel, B. E.~J. 1989, RMxAA, 18, 161

\bibitem[{Pilkington {et~al.}(2012)Pilkington, Gibson, Brook, Calura, Stinson,
  Thacker, Michel-Dansac, Bailin, Couchman, Wadsley, Quinn, \&
  Maccio}]{Pilkington2012}
Pilkington, K., Gibson, B.~K., Brook, C.~B., {et~al.} 2012, Monthly Notices of
  the Royal Astronomical Society, 425, 969

\bibitem[{Proga \& Kallman(2004)}]{2004ApJ...616..688P}
Proga, D., \& Kallman, T.~R. 2004, The Astrophysical Journal, 616, 688

\bibitem[{Proga {et~al.}(2000)Proga, Stone, \& Kallman}]{2000ApJ...543..686P}
Proga, D., Stone, J.~M., \& Kallman, T.~R. 2000, The Astrophysical Journal,
  543, 686

\bibitem[{Read \& Hayfield(2012)}]{2012MNRAS.422.3037R}
Read, J.~I., \& Hayfield, T. 2012, Monthly Notices of the Royal Astronomical
  Society, 422, 3037

\bibitem[{Rejkuba {et~al.}(2005)Rejkuba, Greggio, Harris, Harris, \&
  Peng}]{Rejkuba2005}
Rejkuba, M., Greggio, L., Harris, W.~E., Harris, G. L.~H., \& Peng, E.~W. 2005,
  The Astrophysical Journal, 631, 262

\bibitem[{Rejkuba {et~al.}(2021)Rejkuba, Harris, Greggio, Crnojevi{\'{c}}, \&
  Harris}]{Rejkuba2021}
Rejkuba, M., Harris, W.~E., Greggio, L., Crnojevi{\'{c}}, D., \& Harris, G.
  L.~H. 2021, arXiv, arXiv:2111.02945

\bibitem[{Rejkuba {et~al.}(2014)Rejkuba, Harris, Greggio, Harris, Jerjen, \&
  Gonzalez}]{Rejkuba2014}
Rejkuba, M., Harris, W.~E., Greggio, L., {et~al.} 2014, Astrophysical Journal
  Letters, 791, 2

\bibitem[{Rejkuba {et~al.}(2010)Rejkuba, Harris, Greggio, \&
  Harris}]{Rejkuba2010}
Rejkuba, M., Harris, W.~E., Greggio, L., \& Harris, G. L.~H. 2010, Astronomy
  and Astrophysics, 526, 123

\bibitem[{Rich(1988)}]{Rich1988}
Rich, R.~M. 1988, The Astronomical Journal, 95, 828

\bibitem[{Ritchie \& Thomas(2001)}]{2001MNRAS.323..743R}
Ritchie, B.~W., \& Thomas, P.~A. 2001, Monthly Notices of the Royal
  Astronomical Society, 323, 743

\bibitem[{Rodriguez-Gomez {et~al.}(2016)Rodriguez-Gomez, Pillepich, Sales,
  Genel, Vogelsberger, Zhu, Wellons, Nelson, Torrey, Springel, Ma, \&
  Hernquist}]{Rodriguez-Gomez2016}
Rodriguez-Gomez, V., Pillepich, A., Sales, L.~V., {et~al.} 2016, Monthly
  Notices of the Royal Astronomical Society, 458, 2371

\bibitem[{Rojas-Arriagada {et~al.}(2020)Rojas-Arriagada, Zasowski, Schultheis,
  Zoccali, Hasselquist, Chiappini, Cohen, Cunha, Fern{\'{a}}ndez-Trincado,
  Fragkoudi, Garc{\'{i}}a-Hern{\'{a}}ndez, Geisler, Gran, Lian, Majewski,
  Minniti, Monachesi, Nitschelm, \& Queiroz}]{Rojas-Arriagada2020}
Rojas-Arriagada, A., Zasowski, G., Schultheis, M., {et~al.} 2020, Monthly
  Notices of the Royal Astronomical Society, 499, 1037

\bibitem[{Romano \& Starkenburg(2013)}]{Romano2013}
Romano, D., \& Starkenburg, E. 2013, Monthly Notices of the Royal Astronomical
  Society, 434, 471

\bibitem[{Sazonov {et~al.}(2005)Sazonov, Ostriker, Ciotti, \&
  Sunyaev}]{2005MNRAS.358..168S}
Sazonov, S.~Y., Ostriker, J.~P., Ciotti, L., \& Sunyaev, R.~A. 2005, Monthly
  Notices of the Royal Astronomical Society, 358, 168

\bibitem[{Sazonov {et~al.}(2004)Sazonov, Ostriker, \&
  Sunyaev}]{2004MNRAS.347..144S}
Sazonov, S.~Y., Ostriker, J.~P., \& Sunyaev, R.~A. 2004, Monthly Notices of the
  Royal Astronomical Society, 347, 144

\bibitem[{Schaye {et~al.}(2015)Schaye, Crain, Bower, Furlong, Schaller, Theuns,
  {Dalla Vecchia}, Frenk, McCarthy, Helly, Jenkins, Rosas-Guevara, White, Baes,
  Booth, Camps, Navarro, Qu, Rahmati, Sawala, Thomas, \& Trayford}]{schaye2015}
Schaye, J., Crain, R.~A., Bower, R.~G., {et~al.} 2015, Monthly Notices of the
  Royal Astronomical Society, 446, 521

\bibitem[{Schreiber {et~al.}(2014)Schreiber, Greggio, Falomo, Fantinel, \&
  Uslenghi}]{Schreiber2014}
Schreiber, L., Greggio, L., Falomo, R., Fantinel, D., \& Uslenghi, M. 2014,
  Monthly Notices of the Royal Astronomical Society, 437, 2966

\bibitem[{Soria {et~al.}(1996)Soria, Mould, Watson, {Gallagher, John S.},
  Ballester, Burrows, Casertano, Clarke, Crisp, Griffiths, Hester, Hoessel,
  Holtzman, Scowen, Stapelfeldt, Trauger, Westphal, Soria, Mould, Watson,
  {Gallagher, John S.}, Ballester, Burrows, Casertano, Clarke, Crisp,
  Griffiths, Hester, Hoessel, Holtzman, Scowen, Stapelfeldt, Trauger, \&
  Westphal}]{Soria1996}
Soria, R., Mould, J.~R., Watson, A.~M., {et~al.} 1996, The Astrophysical
  Journal, 465, 79

\bibitem[{Spergel {et~al.}(2007)Spergel, Bean, Dore, Nolta, Bennett, Dunkley,
  Hinshaw, Jarosik, Komatsu, Page, Peiris, Verde, Halpern, Hill, Kogut, Limon,
  Meyer, Odegard, Tucker, Weiland, Wollack, \& Wright}]{2007ApJS..170..377S}
Spergel, D.~N., Bean, R., Dore, O., {et~al.} 2007, The Astrophysical Journal
  Supplement Series, 170, 377

\bibitem[{Springel(2005)}]{2005MNRAS.364.1105S}
Springel, V. 2005, Monthly Notices of the Royal Astronomical Society, 364, 1105

\bibitem[{Strom \& Strom(1978)}]{Strom1978}
Strom, K.~M., \& Strom, S.~E. 1978, The Astrophysical Journal, 83, 73

\bibitem[{Suh {et~al.}(2010)Suh, Jeong, Oh, Yi, Ferreras, \&
  Schawinski}]{Suh2010}
Suh, H., Jeong, H., Oh, K., {et~al.} 2010, The Astrophysical Journal
  Supplement, 187, 374

\bibitem[{Tinsley(1980)}]{Tinsley1980}
Tinsley, B.~M. 1980, Fundamentals of Cosmic Physics, Volume 5, pp. 287-388., 5,
  287

\bibitem[{Toomre {et~al.}(1972)Toomre, Toomre, Toomre, \& Toomre}]{Toomre1972}
Toomre, A., Toomre, J., Toomre, A., \& Toomre, J. 1972, The Astrophysical
  Journal, 178, 623

\bibitem[{Tortora {et~al.}(2010)Tortora, Napolitano, Cardone, Capaccioli,
  Jetzer, \& Molinaro}]{Tortora2010}
Tortora, C., Napolitano, N.~R., Cardone, V.~F., {et~al.} 2010, Monthly Notices
  of the Royal Astronomical Society, 407, 144

\bibitem[{Toyouchi \& Chiba(2018)}]{Toyouchi2018}
Toyouchi, D., \& Chiba, M. 2018, The Astrophysical Journal, 855, 104

\bibitem[{Tsujimoto(2011)}]{Tsujimoto2011}
Tsujimoto, T. 2011, The Astrophysical Journal, 736, 113

\bibitem[{Valenti {et~al.}(2006)Valenti, Ferraro, \& Origlia}]{Valenti2006}
Valenti, E., Ferraro, F.~R., \& Origlia, L. 2006, The Astronomical Journal,
  133, 1287

\bibitem[{van~den Bergh(1962)}]{VandenBergh1962}
van~den Bergh, S. 1962, The Astronomical Journal, 67, 486

\bibitem[{van~den Bergh(1976)}]{VandenBergh1976}
---. 1976, The Astrophysical Journal, 208, 673

\bibitem[{van Dokkum {et~al.}(2014)van Dokkum, Bezanson, van~der Wel, Nelson,
  Momcheva, Skelton, Whitaker, Brammer, Conroy, {F{\"{o}}rster Schreiber},
  Fumagalli, Kriek, Labb{\'{e}}, Leja, Marchesini, Muzzin, Oesch, \&
  Wuyts}]{VanDokkum2014}
van Dokkum, P.~G., Bezanson, R., van~der Wel, A., {et~al.} 2014, The
  Astrophysical Journal, 791, 45

\bibitem[{van Dokkum {et~al.}(2015)van Dokkum, Nelson, Franx, Oesch, Momcheva,
  Brammer, {F{\"{o}}rster Schreiber}, Skelton, Whitaker, van~der Wel, Bezanson,
  Fumagalli, Illingworth, Kriek, Leja, \& Wuyts}]{VanDokkum2015}
van Dokkum, P.~G., Nelson, E.~J., Franx, M., {et~al.} 2015, The Astrophysical
  Journal, 813, 23

\bibitem[{Wang {et~al.}(2020)Wang, Hammer, Rejkuba, Crnojevi{\'{c}}, \&
  Yang}]{Wang2020}
Wang, J., Hammer, F., Rejkuba, M., Crnojevi{\'{c}}, D., \& Yang, Y. 2020,
  Monthly Notices of the Royal Astronomical Society, 498, 2766

\bibitem[{Woodley(2006)}]{Woodley2006}
Woodley, K.~A. 2006, The Astronomical Journal, 132, 2424

\bibitem[{Woosley \& Weaver(1995)}]{1995ApJS..101..181W}
Woosley, S.~E., \& Weaver, T.~A. 1995, The Astrophysical Journal Supplement
  Series, 101, 181

\bibitem[{Yi {et~al.}(2001)Yi, Demarque, Kim, Lee, Ree, Lejeune, \&
  Barnes}]{Yi2001}
Yi, S., Demarque, P., Kim, Y., {et~al.} 2001, The Astrophysical Journal
  Supplement Series, 136, 417

\bibitem[{Zibetti {et~al.}(2020)Zibetti, Gallazzi, Hirschmann, Consolandi,
  Falc{\'{o}}n-Barroso, van~de Ven, \& Lyubenova}]{Zibetti2020}
Zibetti, S., Gallazzi, A.~R., Hirschmann, M., {et~al.} 2020, Monthly Notices of
  the Royal Astronomical Society, 491, 3562

\bibitem[{Zibetti {et~al.}(2005)Zibetti, White, Schneider, \&
  Brinkmann}]{Zibetti2005}
Zibetti, S., White, S. D.~M., Schneider, D.~P., \& Brinkmann, J. 2005, Monthly
  Notices of the Royal Astronomical Society, 358, 949

\bibitem[{Zoccali {et~al.}(2008)Zoccali, Hill, Lecureur, Barbuy, Renzini,
  Minniti, G{\'{o}}mez, \& Ortolani}]{Zoccali2008}
Zoccali, M., Hill, V., Lecureur, A., {et~al.} 2008, Astronomy and Astrophysics,
  486, 177

\bibitem[{Zoccali {et~al.}(2017)Zoccali, Vasquez, Gonzalez, Valenti,
  Rojas-Arriagada, Minniti, Rejkuba, Minniti, McWilliam, Babusiaux, Hill, \&
  Renzini}]{Zoccali2017}
Zoccali, M., Vasquez, S., Gonzalez, O.~A., {et~al.} 2017, Astronomy and
  Astrophysics, 599, 12

\end{thebibliography}
